\documentclass{amsart}%siamltex1213}
\usepackage{hyperref}
\usepackage{graphicx}
\usepackage{amsfonts}
\usepackage{amsmath}
\usepackage[usenames,dvipsnames]{xcolor}
\usepackage{soul}

\DeclareGraphicsExtensions{.pdf,.png}

\newcommand{\eps}{\varepsilon}

\newcommand{\beq}{\begin{equation}}
\newcommand{\eeq}{\end{equation}}

\newcommand{\ubar}{{\overline{u}}}
\newcommand{\cc}{{\mathrm{c.c.}}}
\def\barr{\begin{array}}
\def\earr{\end{array}}

\title{Nonlinear Schr\"odinger equations \\ and the universal
  description \\ of dispersive shock wave structure 
   }

\author{T.~Congy$^\textrm{\textdagger}$, G.~A. El$^\textrm{\textdagger}$}
\thanks{$^\textrm{\textdagger}$Department of Mathematics, Physics and Electrical Engineering, Northumbria 
  University, Newcastle upon Tyne, UK}
\author{M.~A. Hoefer$^\textrm{\textdaggerdbl}$}
\thanks{$^\textrm{\textdaggerdbl}$Department of Applied Mathematics,
  University of 
  Colorado, Boulder, USA}
\author{M. Shearer$^\textrm{\textsection}$}
\thanks{$^\textrm{\textsection}$Department of Mathematics, North Carolina State
  University, Raleigh,  USA}

\begin{document}
\maketitle

\pagestyle{myheadings}
\thispagestyle{plain}
\markboth{T.~CONGY, G.~A. EL, M.~A. HOEFER, AND M.~ SHEARER}{NLS EQUATION AND THE UNIVERSAL DESCRIPTION OF DSW STRUCTURE} 

\begin{abstract}
The nonlinear Schr\"odinger (NLS) equation and the Whitham
  modulation equations both describe slowly varying, locally periodic
  nonlinear wavetrains, albeit in differing amplitude-frequency
  domains.  In this paper, we take advantage of the overlapping
  asymptotic regime that applies to both the NLS and Whitham
  modulation descriptions in order to develop a universal analytical
  description of dispersive shock waves (DSWs) generated in Riemann
  problems for a broad class of integrable and non-integrable
  nonlinear dispersive equations.  The proposed method extends DSW
  fitting theory that prescribes the motion of a DSW's edges into the
  DSW's interior, i.e., this work reveals the DSW
  structure. Our approach also provides a natural framework in which to analyze
  DSW stability. We consider several representative, physically
  relevant examples that illustrate the efficacy of the developed
  general theory.  Comparisons with direct numerical simulations show
  that inclusion of higher order terms in the NLS equation enables a
  remarkably accurate description of the DSW structure in a broad region that
extends from the harmonic, small amplitude edge.
  \end{abstract}

\section{Introduction}
\label{sec:introduction}

There has been a surge of interest recently in the subject of dispersive
hydrodynamics and, in particular,  dispersive shock waves (DSWs) 
(see \cite{biondini_dispersive_2016, el_dispersive_2016}
and references therein). This has largely occurred thanks to the
growing recognition of the fundamental nature and ubiquity of DSWs in
physical applications: from shoaling tsunami waves
\cite{tissier_nearshore_2011, arcas_seismically_2012,
  grimshaw_depression_2016} and internal undular bores in the ocean
\cite{smyth_hydraulic_1988, scotti_observation_2004,
  grimshaw_internal_2018} and atmosphere \cite{christie_morning_1992,
  porter_modeling_2002} to nonlinear diffraction patterns and optical
shocks in laser beam propagation \cite{wan_dispersive_2007,
  ghofraniha_shocks_2007, conti_observation_2009,
  fatome_observation_2014,xu_dispersive_2017-1}, quantum shocks in
superfluids \cite{dutton_observation_2001, hoefer_dispersive_2006-1,
  rolley_hydraulic_2007,hoefer_matter-wave_2009}, and nonlinear spin
wave propagation in magnetic thin films
\cite{janantha_observation_2017}. On the other hand, the study of DSWs
has revealed a number of challenging mathematical problems in the
context of both integrable and non-integrable nonlinear wave
equations.

\begin{figure}
\centering
\includegraphics[width=0.5\textwidth]{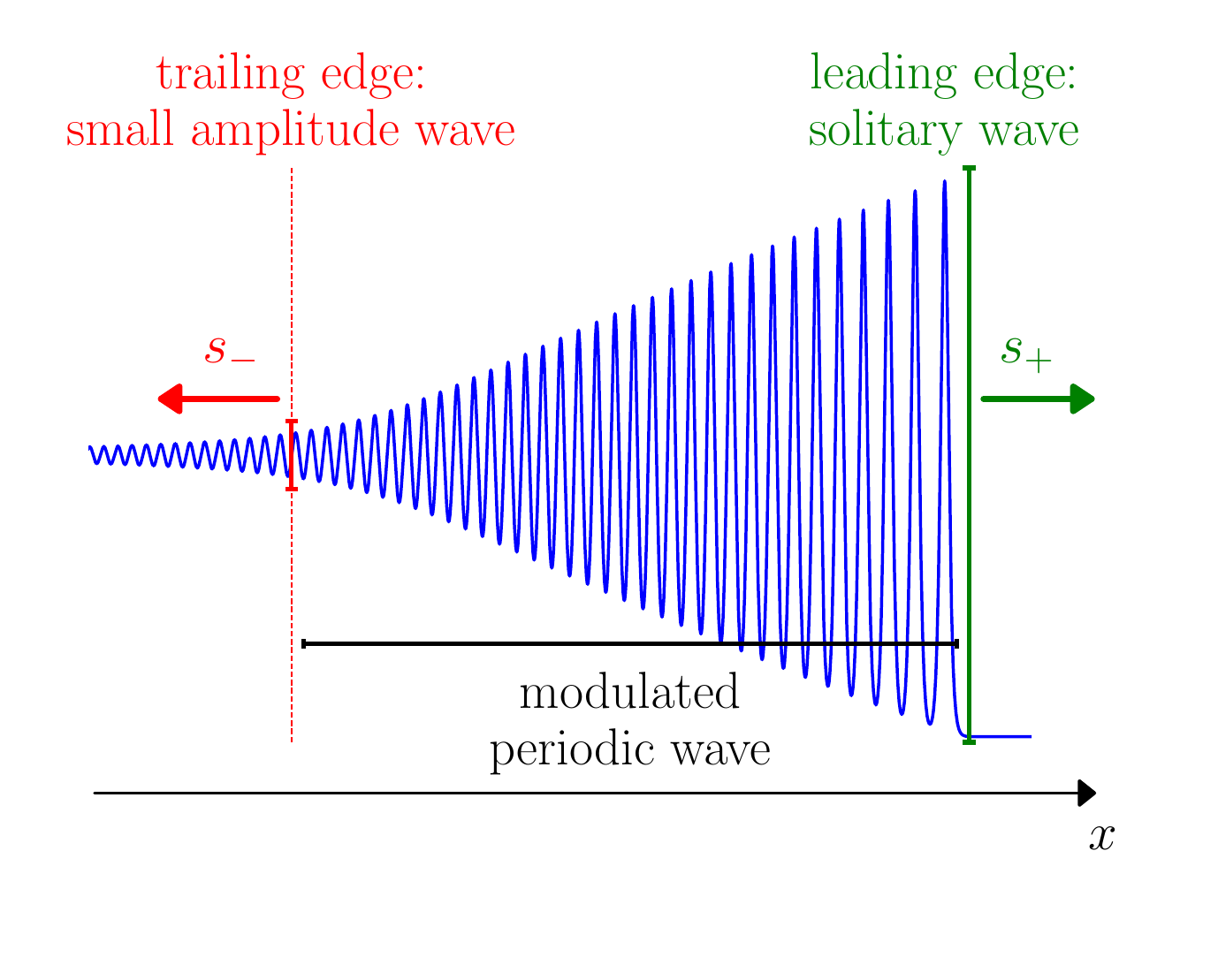}
\caption{DSW expanding oscillatory structure in convex dispersive hydrodynamics with
  negative dispersion.}
\label{fig:Fig1}
\end{figure}

A DSW is an expanding, modulated nonlinear wavetrain that connects two
disparate hydrodynamic states (see Fig.~\ref{fig:Fig1}). It can be viewed as a dispersive
counterpart to the dissipative, classical shock.  Hydrodynamic wave
breaking singularities in dispersive media are generically resolved by
DSWs.  A DSW has a distinct multi-scale structure consisting of an
oscillatory transition between two non-oscillatory---e.g., slowly
varying or constant---states: one edge is associated with a solitary wave
or soliton (for convenience, we use the term soliton regardless of the
integrability of the governing equation) that is connected, via a slowly
modulated periodic wavetrain, to a harmonic, small-amplitude wave at
the opposite edge. The relative position
(left/trailing or right/leading) of the soliton and harmonic edges
determines the DSW orientation $d$, found in terms of the curvature  of the
linear dispersion relation as $d=-\hbox{sgn}[\partial_{kk}\omega_0(k,
u_0)]$ \cite{el_dispersive_2017, el_dispersive_2016}. Here, $\omega=
\omega_0(k, u_0)$ is the frequency of a small amplitude wave with
wavenumber $k$ that propagates on the mean flow background $u_0$.  The
DSW shown in Fig.~\ref{fig:Fig1} has $d=1$ because the solitary wave
is on the rightmost, leading edge.  The shock structure of a DSW---an
unsteady oscillatory wavetrain---is more complex than the
stationary shock structure of a viscous shock wave.  In particular, a DSW
cannot be described by a traveling wave (ODE) solution of the nonlinear wave
equation
\cite{el_dispersive_2017}.

The rapidly oscillating structure of DSWs motivates the use of
asymptotic, WKB-type, methods for its analytic description. One such
method, known as Whitham modulation theory
\cite{whitham_non-linear_1965, whitham_linear_1974} (see also
 \cite{ kamchatnov_nonlinear_2000}),  is based on the averaging of dispersive
hydrodynamic conservation laws over nonlinear periodic wavetrains  leading to a system of first order quasilinear
partial differential equations (PDEs).
%Because it is agnostic to more subtle---and in practice,
%rare---mathematical properties of the governing nonlinear wave
%equation, 
Whitham theory has proved particularly effective for the
description of DSWs in both integrable and non-integrable systems.   If the dispersive
hydrodynamics are described by an integrable equation such as the
Korteweg-de Vries (KdV) or nonlinear Schr\"odinger (NLS) equation, the
associated Whitham system can be represented in a diagonal, Riemann
invariant form \cite{whitham_non-linear_1965,
  flaschka_multiphase_1980, kamchatnov_nonlinear_2000}. This fact
enabled Gurevich and Pitaevskii (GP)
\cite{gurevich_nonstationary_1974} to construct an explicit modulation
solution for a DSW generated by a Riemann problem for the KdV
equation. The GP construction is based on a self-similar, rarefaction wave solution of
the KdV-Whitham equations.  This modulation solution describes the
interior shock structure of a DSW and reveals a monotonic change in
the nonlinear wave's wavenumber, mean, and amplitude as the DSW is
spatially traversed.

% The existence of Riemann invariants for a modulation system is a
% consequence of complete integrability, via the inverse scattering
% transform, of the original, dispersive equation
% \cite{flaschka_multiphase_1980} and is crucial for the obtaining the
% full effective analytical description of a DSW. Indeed, Riemann
% invariants and the DSW modulation solutions have been found for
% other integrable equations such as the modified KdV equation,
% defocusing nonlinear Schr\"odinger (NLS) equation and oth. (see
% \cite{el_dispersive_2016} and references therein).

For non-integrable dispersive equations, diagonalization of the
associated modulation systems in terms of Riemann invariants is
generally not possible, often presenting an insurmountable obstacle to
the explicit determination of the Whitham system's simple wave
solution, although its existence requires only strict hyperbolicity
and genuine nonlinearity.  Consequently, the analytical description of
a DSW's interior structure has so far been limited to integrable
systems or a detailed analysis of the Whitham modulation system in
certain limiting regimes on a case-by-case basis
\cite{gurevich_nonlinear_1990,el_unsteady_2006}.  One can, however,
explicitly determine key observables associated to each DSW edge, even
for non-integrable equations.  These observables include the DSW edge
speeds and their associated wave parameters---the harmonic edge
wavenumber and the soliton edge amplitude.  The determination of these
observables represents the fitting of a DSW to the long-time dynamics
of piecewise constant, initial Riemann data.  The DSW fitting method
proposed in \cite{el_resolution_2005} (see also
\cite{el_dispersive_2016}) is based on a fundamental, generic
property: the Whitham modulation equations admit exact reductions to a
set of common, much simpler, analytically tractable equations in the
limits of vanishing amplitude and vanishing wavenumber, which
correspond to the harmonic and soliton DSW extremes (see
Fig.~\ref{fig:Fig1}). Therefore, the DSW fitting method can be viewed
as a universal dispersive hydrodynamic analog to the Rankine-Hugoniot
conditions for dissipative, viscous shocks. The key advantage of the
method is that it  involves neither the determination nor the analysis of
the full Whitham system because the required zero-amplitude and
zero-wavenumber reductions are available directly and are ultimately
determined by the nonlinear, hyperbolic flux and the linear dispersion
relation of the dispersive hydrodynamics. The method greatly expands
the scope of DSW analysis as it is not reliant on integrability of the
governing nonlinear dispersive equation via inverse scattering
theory. It has been successfully applied to many non-integrable
dispersive hydrodynamic systems.  See, for example,
\cite{el_unsteady_2006, el_theory_2007, esler_dispersive_2011,
  crosta_whitham_2012,lowman_dispersive_2013,hoefer_shock_2014,
  maiden_observation_2016,hoefer_oblique_2017,an_dispersive_2018}. Restrictions to the
method's applicability are related to possible violations of genuine
nonlinearity (convexity) and strict hyperbolicity of the modulation
system \cite{lowman_dispersive_2013,hoefer_shock_2014}.

Once the parameters of the leading and trailing edges have been
determined by DSW fitting, wave modulation in the  vicinity
of these edges can be, in principle, determined by expanding the full
Whitham system, for small amplitudes near the harmonic edge and small
wavenumbers near the soliton edge \cite{gurevich_nonlinear_1990,
  el_unsteady_2006}.  Such an asymptotic consideration, however, has a
number of serious drawbacks due to the need to derive and analyze
the full modulation system. Apart from being a potentially
daunting technical task, this presents a major disadvantage to the whole
procedure as it is system specific.  It would
therefore be highly desirable to have a more direct, widely applicable
method for the determination of the DSW structure including modulation
near the harmonic and soliton edges, which would complement and extend
the existing general DSW fitting procedure.

In this paper, we develop a general method for the determination of
the universal nonlinear DSW modulation---the DSW structure---near the
harmonic edge. This asymptotic modulation provides crucial information
about the variation of the amplitude in the DSW (i.e. the envelope of the oscillatory wavetrain)  
as well as other
physical DSW parameters such as the wavenumber and mean flow.  The modulation is
universal because it is derived from the NLS equation,  a universal model of weakly nonlinear, modulated dispersive
wavetrains \cite{benney_propagation_1967}.  The present work takes
advantage of the asymptotic overlap region in the vicinity of the DSW
harmonic edge between the semiclassical, long-wave limit of the NLS
equation and the small amplitude limit of the full Whitham modulation
equations.  The commonalities and differences between Whitham
modulation theory and the NLS equation have been widely discussed in
the literature (see, e.g., \cite{newell_solitons_1985,
  el_dispersive_2016}) but to the best of our knowledge, the overlap
regime for the applicability of the two descriptions has never been
used in practice, except very recently in
\cite{maiden_modulations_2016}.  While the Whitham equations describe
modulations of fully nonlinear wavetrains, the NLS description is
advantageous in the weakly nonlinear regime because it incorporates
higher order dispersive effects of the wave envelope that are not
accounted for in leading order Whitham theory.

We use the parameters obtained by DSW fitting applied to the harmonic
edge as input information for a standard, small amplitude, multiple
scales expansion that leads to the NLS equation for the wave's
envelope and phase.  The specific information relevant to dispersive
hydrodynamics consists of the NLS' nonlinear and dispersion coefficients.  The
universal asymptotic modulation in a DSW is then found as a special
\textit{vacuum rarefaction} simple wave solution of the NLS equation
in the long-wavelength, dispersionless limit, which is similar to the solution to the shallow water equations for the classical dam
break problem with a dry downstream bed.  

We consider several
representative integrable and non-integrable examples to illustrate
the efficacy of the developed general theory. Comparisons with direct
numerical simulations show that the accuracy of the asymptotic
description improves dramatically when higher order terms of the NLS
equation are taken into account in the so-called HNLS equation.  The
HNLS equation was first derived in the nonlinear optics context
\cite{kodama_optical_1985, kodama_nonlinear_1987} but it is a
universal equation that also arises in other applications including
fluid dynamics \cite{grimshaw_long-time_2008,
  sedletsky_fourth-order_2003} and plasma physics
\cite{gromov_nonlinear_1996}.
We observe that in all considered examples, the vacuum rarefaction
simple wave solution of the semi-classical, dispersionless HNLS
equation provides a remarkably accurate description of the DSW
modulation, and therefore the DSW structure, in a broad vicinity of
the harmonic edge. Finally, we show that convexity of the linear
dispersion relation for the original dispersive hydrodynamics plays a
key role in the determination of DSW stability via the
focusing/defocusing character of the associated NLS equation.  

The paper is organized as follows. We begin in
Sec.~\ref{sec:dsw_mod_theory} with a brief outline of the necessary
elements of DSW modulation theory and, in particular, the DSW fitting
method for the determination of the DSW harmonic edge in scalar
dispersive hydrodynamic systems. Section \ref{sec:small_ampl_dsw}
develops an asymptotic, multiple scales expansion in the vicinity of
the DSW harmonic edge that leads to the NLS equation.  This is used to
derive the universal, first order approximation of the DSW modulation
as a vacuum rarefaction simple wave solution of the long-wave,
dispersionless limit of the NLS equation. We then extend the first
order analysis by including higher order terms in the multiple scales
expansions in Sec.~\ref{sec:higher_order_nls}.  This results in the
HNLS equation for which we find the appropriate simple wave solution
in the long-wavelength limit.  Section \ref{sec:nls_examples}
is devoted to applications to several
representative examples.  The examples include the KdV equation, the conduit equation that models the
interfacial dynamics of a rising, buoyant, viscous fluid within
another miscible, high viscosity contrast fluid
\cite{scott_observations_1986,lowman_dispersive_2013}, and the
Serre equations for fully nonlinear shallow water waves
\cite{serre_contribution_1953, su_kortewegvries_1969,
  el_unsteady_2006}.  The latter two
equations are non-integrable.  We complete the paper with a summary,
and discussion of further directions in Sec.~\ref{sec:discussion}.
Appendices A and B detail the multiple scales derivations of the NLS
and HNLS equations for  the conduit equation and the Serre system.
% Both derivations have been performed using symbolic computations in
% {\it Mathematica}.
Appendix C is devoted to a brief description of
numerical methods used for simulations.

\section{Dispersive shock waves: modulation theory}
\label{sec:dsw_mod_theory}

In this Section, we outline the elements of DSW modulation theory that
are necessary for developments in subsequent sections.  A detailed
exposition can be found in \cite{el_dispersive_2016,el_dispersive_2017}.

\subsection{Modulation equations and the matching regularization of
  the Riemann problem}

We consider scalar dispersive hydrodynamics generically described by
the equation
\begin{equation}
  \label{uni_dh}
  u_t+ f(u)_x+D[u]_x=0
\end{equation}
with $f''(u) \ne 0$.  The dispersive operator $D$  (generally
integro-differential) is assumed to have the property that equation (\ref{uni_dh}) admits the
real-valued linear dispersion relation $\omega=\omega_0(k, u_0)$ with
long-wave expansion
\begin{equation}
  \label{eq:lin_dr2}
  \omega_0(k, u_0)=f'(u_0) k + \zeta k^3 +o(k^3), \quad k \ll 1, \quad
  \zeta \ne 0
\end{equation}
for small-amplitude waves proportional to $e^{i(kx - \omega t)}$ and
propagating on a constant (or slowly varying) background $u=u_0$.  We
shall initially assume that the dispersion relation is purely convex
or concave, so that $\partial_{kk}\omega_0 \ne 0$. Suppose equation
\eqref{uni_dh} has a three-parameter periodic traveling wave solution
and at least two local conservation laws. These basic requirements are
quite generic and are satisfied by many dispersive hydrodynamic
equations arising in applications. When the dispersive hydrodynamics
admit the above properties, we say that they are of KdV type.

\begin{figure}
  \centering
  \includegraphics[width=8cm]{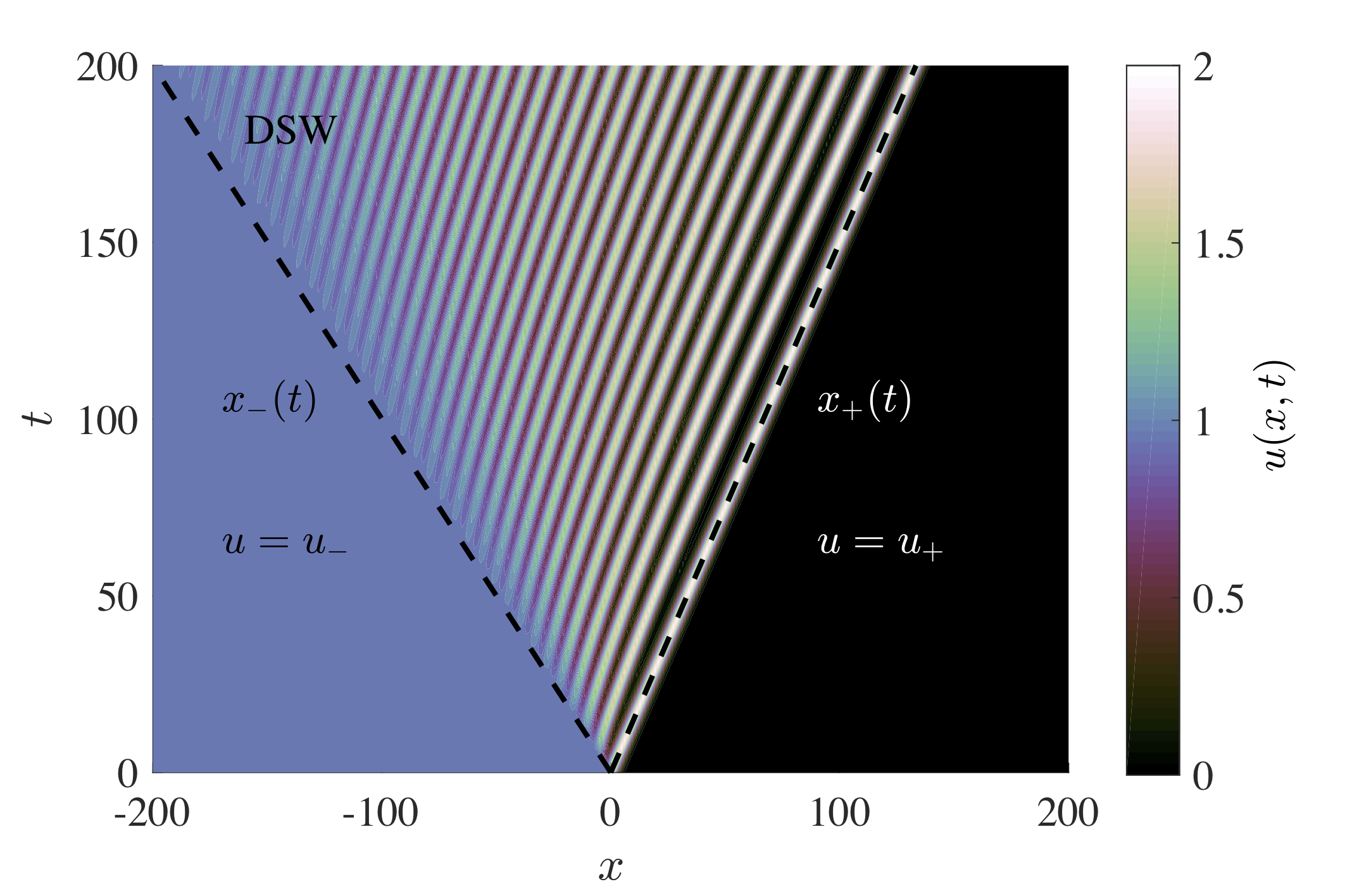}
  \caption{Contour plot of the asymptotic KdV DSW solution obtained by
    GP \cite{gurevich_nonstationary_1974} for $u_- = 1$, $u_+ = 0$.}
  \label{fig:kdv_dsw_contour}
\end{figure}

We shall consider the evolution of Riemann step initial data
\begin{equation}
  \label{eq:riemann_data1}
  u(x,0)=\left\{
    \begin{array}{ll}
      u_-, \quad & x<0\\[6pt]
      u_+, &x>0
    \end{array}
  \right.
\end{equation} 
for Eq.~(\ref{uni_dh}).  Our consideration will be based upon the
fundamental assumption that the initial step \eqref{eq:riemann_data1}
is regularized in the long-time limit by the emergence of three
distinct regions in the $x$-$t$ upper half space-time plane so that
the solution is given by two constant states $u=u_-$ and $u=u_+$ that
are separated by an expanding DSW region (see
Fig.~\ref{fig:kdv_dsw_contour}).  Within this region, the solution has
an oscillatory structure described by a modulated, locally periodic
wavetrain that exhibits a solitary wave at one edge and a
vanishing amplitude linear wave at the opposite edge (recall
Fig.~\ref{fig:Fig1}). This asymptotic structure of the Riemann problem
solution has been rigorously recovered for a number of integrable
equations (see, e.g.,
\cite{egorova_long-time_2013,jenkins_regularization_2015}). For
non-integrable equations such as the Serre system
\cite{el_unsteady_2006} or conduit equation
\cite{lowman_dispersive_2013}, the existence of a modulated, periodic,
single phase wave structure of a DSW is a plausible assumption which
can be inferred from numerical simulations.
 
We assume that Eq.~\eqref{uni_dh} admits a three parameter family of
periodic, traveling wave solutions $u(x,t) = U(\theta)$, where $\theta=kx-\omega t$,
$k $ being the wavenumber and $\omega$  the wave
frequency, so that $U(\theta+ 2\pi) = U(\theta)$. It is convenient to
use the period mean $\ubar=(2\pi)^{-1}\oint U {\rm d} \theta$, the
amplitude $a=u_{\rm max} - u_{\rm min}$ and the wavenumber $k$ as a
basic parameter set, i.e., $U(\theta) \equiv U(\theta; \ubar, a, k)$;
all other physical parameters, such as the frequency $\omega$ or the
mean square $\overline{u^2}$ are functions 
of the basic triple $(\ubar,a,k)$. We also assume that the solution
$U(\theta; \ubar, a, k)$ has two asymptotic limits: (i) when $a \to 0$
it transforms into a linear wave on the background $u=u_0$ with the
dispersion relation $\omega=\omega_0(k, u_0)$; (ii) when $k \to 0$ it
transforms into an exponentially decaying solitary wave.  Examples of
dispersive equations whose periodic solutions satisfy the above
properties include KdV, modified KdV, the conduit equation and others.

We now consider a solution of Eq.~\eqref{uni_dh} represented by the $2\pi$-periodic traveling
wave  with slow $(x,t)$-dependence of $(\ubar, a, k)$. This slowly varying traveling wave is characterised by the generalized phase $\theta(x,t)$ so that the local wavenumber and frequency are given by $k=\theta_x$ and $\omega = -\theta_t$ respectively. The variations of $(\ubar, a, k)$ satisfy the Whitham
modulation equations \cite{whitham_linear_1974}, which can be obtained
by applying multiple scales expansions or, equivalently, by averaging
two independent conservation laws of \eqref{uni_dh} over the periodic
family and completing the system with the consistency equation
$\theta_{xt}=\theta_{tx}$ that yields wave number conservation
$k_t + \omega_x=0$. The same set of modulation equations can be
derived via an averaged variational principle
\cite{whitham_two-timing_1970}.  Assuming non-vanishing Jacobians, the
Whitham system can be represented as a system of quasilinear first
order equations
\begin{equation}
  \label{whitham_scal}
  \left(
    \begin{array}{c}
      \ubar \\ a \\ k \\
    \end{array}
  \right)_t +
  \mathrm{A}(\ubar, a, k) 
  \left(
    \begin{array}{c}
      \ubar \\ a \\ k\\
    \end{array}
  \right)_x =
  \left(
    \begin{array}{c}
      0 \\ 0 \\ 0\\
    \end{array}
  \right),
\end{equation}
where $\mathrm{A}(\ubar, a, k) \in \mathbb{R}^{3 \times 3}$ is a
matrix. We initially assume hyperbolicity so that the eigenvalues
$V_1 \le V_2 \le V_3$ of $ \mathrm{A}$ are real and the eigenvectors
$\{{\bf r}_j ~|~ \mathrm{A} {\bf r}_j =V_j {\bf r_j}, ~ j=1,2,3 \}$
form a basis in $\mathbb{R}^{3}$.

In the context of a DSW that is described by a modulated periodic wave
solution, the Whitham equations \eqref{whitham_scal} are subject to
free boundary (matching) conditions
\cite{gurevich_nonstationary_1974,el_resolution_2005}
\begin{equation}
  \label{gp_matching}
  \begin{split}
    x=x_-(t): \qquad a=0, \ \ \ubar = u_- \, ,\\
    x=x_+(t): \qquad k=0, \ \ \ubar = u_+ \, ,
  \end{split}
\end{equation}
thus ensuring continuity of the mean flow $\ubar$ at the unknown DSW
edges $x=x_\pm(t)$. Outside the DSW region $x_-(t)\le x \le x_+(t)$,
the solution is given by $u=u_-$ for $x < x_-(t)$ and $u=u_+$ for
$x>x_+(t)$. Here, for specificity, we have assumed a positive DSW
orientation (see Fig.~ \ref{fig:Fig1}) so that the harmonic edge is trailing, $x=x_-(t)$, and the
soliton edge is leading, $x=x_+(t)$. We also assume concave flux,
$f''(u)>0$, which implies the admissibility or causality condition
$u_->u_+$ for a compressive DSW \cite{el_dispersive_2017, el_dispersive_2016}. The cases
of positive dispersion that yield either a negative DSW orientation, $d=-1$,
or convex flux $f''(u)<0$ (which implies $u_-<u_+$ for compressive
DSW formation) can be considered in a similar fashion.

The hydrodynamic scaling invariance $x \to a x$, $t \to a t$ of both
the modulation equations \eqref{whitham_scal} and initial conditions
\eqref{eq:riemann_data1}, together with hyperbolicity, necessitate a
self-similar, simple wave modulation solution. To satisfy the matching
conditions \eqref{gp_matching}, the modulation solution must be a
2-wave rarefaction curve \cite{el_resolution_2005,el_dispersive_2016}
given by
\begin{equation}\label{2wave}
 V_2(\ubar, a, k)= x/t, \qquad I_1(\ubar, a, k)=0, \quad I_2(\ubar, a, k)=0,
\end{equation}
where $I_{1,2}$ are integrals of the Whitham system
\eqref{whitham_scal} on the solution curve. The integrals are
parametrized by the Riemann data $u_{\pm}$, i.e., $I_1(u_-,0,k_-)
= 0$ and $I_2(u_+,a_+,0) = 0$ determine the trailing edge harmonic
wavenumber $k_-$ and the leading edge soliton amplitude $a_+$.  For
the KdV equation solution, Eq.~\eqref{2wave} was found explicitly by
GP \cite{gurevich_nonstationary_1974} in terms of Riemann invariants
that are available for the KdV-Whitham system owing to its complete
integrability \cite{flaschka_multiphase_1980}, unbeknownst to Whitham
who was able to determine the Riemann invariants explicitly via an
involved, direct calculation \cite{whitham_non-linear_1965} (see also \cite{bhatnagar_nonlinear_1980, kamchatnov_nonlinear_2000}). The DSW
edge speeds $s_{\pm}$ are constant and follow from the modulation
solution \eqref{2wave} in the $a \to 0$ (harmonic, trailing edge) and
$k \to 0$ (soliton, leading edge) limits by evaluating the linear
group velocity and soliton phase velocity, respectively. The
$(x,t)$-contour plot of the GP solution to the Riemann problem for the
KdV equation $u_t+uu_x+u_{xxx}=0$ illustrates the described modulation
theory setting in Fig.~\ref{fig:kdv_dsw_contour}.

The above outlined construction of the DSW modulation solution for
scalar equations \eqref{uni_dh} can be extended to systems describing
bidirectional dispersive hydrodynamics, e.g., the Serre shallow water
equations, the generalized NLS equation and other systems.  See, e.g.,
\cite{el_resolution_2005, el_undular_2005, el_unsteady_2006,
  hoefer_shock_2014}.

\subsection{The determination of the harmonic edge: DSW fitting}
\label{sec:dsw_fitting}

Modulation systems \eqref{whitham_scal} obtained by averaging
dispersive hydrodynamic systems \eqref{uni_dh} exhibit an important
general property: they admit exact reductions to lower order
quasilinear systems in the harmonic ($a \to 0$) and soliton ($k \to
0$) limits (recall that these two limits correspond to special wave regimes realized at the DSW
edges \cite{gurevich_nonlinear_1990, el_resolution_2005,
  el_dispersive_2016}). Importantly, these exact reductions are often
available directly, without the necessity to derive the full
modulation system.  Another fundamental fact is that in the Riemann
problem, the DSW edges are characteristics when the modulation system
\eqref{whitham_scal} is hyperbolic. As a result, the speeds $s_{\pm}$
of the harmonic and soliton edges can be obtained from the analysis of
the reduced modulation systems.  The corresponding technique proposed
in \cite{el_resolution_2005} is sometimes referred to as the DSW
fitting method.

Determining the harmonic edge via the DSW fitting method is
particularly simple. Indeed, the harmonic reduction ($a=0$) of the
modulation system \eqref{whitham_scal} can be shown to be universally
represented in the physically transparent form
\begin{equation}
  \label{lin_red}
  \ubar_t + f(\ubar)_x=0,  \qquad k_t + [\omega_0(k, \ubar)]_x=0.
\end{equation}
Then the DSW harmonic edge speed coincides with the linear group
velocity for the edge parameters $\ubar=u_-$, $k=k_-$:
\begin{equation}
  s_- = \partial_k\omega_0(k_-, u_-) \quad  \hbox{where} \quad  k_-=K(u_-),
\end{equation}
 $K(\ubar)$ being the harmonic edge wavenumber locus function, which
 is determined as follows.  Let the value $\ubar =u_+$ at the DSW
 soliton edge be fixed. Then the function $K(\ubar)$ is found from the
 ODE
\begin{equation}\label{ode_char}
\frac{dK}{d \ubar} = \left[\frac{\partial_{\ubar} \omega_0}
{f'({\ubar}) - \partial_k \omega_0 }\right]_{k=K}, \quad K(u_+)=0.
\end{equation}
The ODE \eqref{ode_char} follows by integrating the differential
associated with the group velocity characteristic of
Eq.~\eqref{lin_red} on the DSW harmonic edge.  It specifies a relation
between admissible values of $k$ and $\ubar$ along this edge.  The
initial condition in \eqref{ode_char} follows from the GP matching
conditions \eqref{gp_matching} (see
\cite{el_resolution_2005,el_dispersive_2016} for details).
%the natural
%requirement that, if $u_- \to u_+$ then the DSW must be vanishing,
%i.e., $x_- \to x_+$. To be consistent with the GP conditions
%\eqref{gp_matching} in that case, one must require $k(x_-,t) \to
%k(x_+,t)=0$, which implies that $K(u_+)=0$ in the limit. 

The determination of the soliton edge is analogous, although it
involves some nontrivial change of variables which we do not describe
here (see \cite{el_resolution_2005, el_dispersive_2016}). The extension of scalar DSW
fitting to bidirectional, Eulerian dispersive hydrodynamic systems has
been developed in \cite{el_resolution_2005,hoefer_shock_2014}.  The
DSW fitting procedure is subject to certain admissibility conditions
derived from the monotonicity requirement for the relevant real
characteristic velocity along the integral curve \eqref{2wave}, i.e.,
the modulation system must be genuinely nonlinear and strictly
hyperbolic along the entire integral curve
\cite{lowman_dispersive_2013,hoefer_shock_2014}.  Therefore, the DSW
fitting construction is not reliant upon the integrability of the
dispersive hydrodynamic evolution equation \eqref{uni_dh}, it only
requires strict hyperbolicity and genuine nonlinearity of the
modulation system \eqref{whitham_scal}.

\section{Small amplitude DSW regime and the NLS equation}
\label{sec:small_ampl_dsw}

We shall be interested in the region of a DSW adjacent to the harmonic
edge $x=s_-t$, where the oscillation amplitude $a$ is relatively
small. One can, in principle, expand the Whitham equations
\eqref{whitham_scal} in powers of $a$ for $a \ll 1$ and solve them by
seeking a solution in powers of the amplitude, subject to the matching
condition \eqref{gp_matching} at $x=s_-t$. This programme has been to
some extent realized in \cite{gurevich_nonlinear_1990,
  el_unsteady_2006} for several non-integrable dispersive hydrodynamic
systems, including the equations for ion-acoustic and magneto-acoustic
waves in collisionless plasma and the Serre equations for fully
nonlinear dispersive shallow water waves. In all cases, the full
modulation system \eqref{whitham_scal} (or its bi-directional
generalization) was derived, and then reduced to an abstract, model Whitham
system for $a$ and $k$ (see \cite{whitham_linear_1974}, Ch.~16.15)
involving an effective nonlinear frequency correction $\omega_2(k,
u_-)a^2$ to the linear dispersion relation $\omega_0(k, u_-)$. As a
result, the first order DSW modulation near the harmonic edge was
determined, including the linear growth of the amplitude with
distance. This approach, however, has a major drawback, due to its reliance 
on the full modulation system in each case while only the small
amplitude expansion is actually used.  
% It would be, therefore, highly desirable to have a more
% direct and universal method for the determination of the small
% amplitude DSW modulation near the harmonic edge.  Such a method would
% complement and extend the DSW fitting procedure described in the
% previous Section. % Redundant with intro

Here,  instead of deriving the full modulation system
and making subsequent small amplitude expansions, we perform an
appropriate small amplitude expansion directly on the original system
and then derive modulation equations for weakly nonlinear periodic
(Stokes) waves \cite{whitham_linear_1974}.  Slow modulations of almost
monochromatic Stokes waves for a broad class of nonlinear dispersive
systems are known to be described by the nonlinear Schr\"odinger
equation or its higher order versions
\cite{benney_propagation_1967,ablowitz_nonlinear_2011,
  yang_nonlinear_2010}).  
% It is this reason that NLS is deemed universal. The commonalities
% and differences between Whitham modulation theory and the NLS
% equation have been widely discussed in the literature, see
% \cite{newell_solitons_1985, el_dispersive_2016}.
% Both provide mathematical description of
% nonlinear modulated waves, however, the Whitham method has no
% amplitude restrictions and incorporates a broad range of
% wavenumbers/frequencies while the NLS equation describes small
% amplitude, nearly monochromatic waves. On the other hand, the Whitham
% equations are quasilinear and don't contain dispersive terms, while
% the NLS equation is capable of capturing dispersive behavior {\it of
%   the wave envelope}.  % Redundant with intro
Consequently both Whitham modulation
theory and an NLS description can be used in the inner vicinity of the DSW harmonic edge.  Moreover, since a DSW is described by a
rarefaction wave solution of the Whitham equations, the counterpart
NLS description will reduce to finding a simple wave solution to the
{\it dispersionless limit} of the corresponding NLS equation or one of
its higher order versions.

%Summarising, instead of constructing the full
%modulation system, then finding its weakly nonlinear approximation,
%and then integrating it, one can take advantage of the direct
%small-amplitude NLS approximation of the dispersive hydrodynamic
%equation \eqref{uni_dh} and use it to get the required asymptotic
%modulation solution.

It is instructive to start with an outline of the standard derivation
of the NLS equation.  See, e.g., \cite{ablowitz_nonlinear_2011} for
examples and further details.  Let $\eps$ be a small parameter
characterizing the wave amplitude.  We seek the solution of the
dispersive hydrodynamic equation \eqref{uni_dh} in the form of an
asymptotic expansion about the constant $u_0$ for a nearly monochromatic wave with the dominant
carrier wavenumber $k$
\begin{equation} 
  \label{exp1}
  u=u_0+\eps u_1+\eps^2 u_2+\eps^3 u_3+\dots ,
\end{equation} where
$$
u_1=A(X,T_1,T_2)e^{i(kx-\omega t)}+\hbox{c.c.}, \ X=\eps x, T_1=\eps
t, T_2=\eps^2 t.
$$
Substituting expansion \eqref{exp1} into Eq.~\eqref{uni_dh} and
collecting terms in powers of $\eps$, we obtain the linear dispersion
relation $\omega=\omega_0(k,u_0)$ at the first order.  To eliminate
secular terms at $O(\eps^2)$, we require that the complex wave
envelope move with the group velocity
\begin{equation}
\label{vg}
A_{T_1}+\partial_k\omega_0 A_X=0.
\end{equation}
The NLS equation arises as the condition for removal of secular terms
at $O(\eps^3)$ and has the form
\begin{equation}
  \label{NLS}
  iA_{T_2} +\beta A_{XX} +\gamma |A|^2 A=0,
\end{equation}
where $\beta(k,u_0)= \tfrac12 \partial_{kk}\omega_0$. We also obtain
the variation of the mean $\ubar - u_0 \sim \eps^2 b_1(k, u_0)|A|^2$
as a by-product of the $O(\eps^3)$ calculation. Here, the factors $b_1$ and $\gamma$
has no general expressions.

Although the outlined above derivation is standard, it can be quite
laborious, especially for systems. The difficult part of the
derivation is the determination of the nonlinear coefficient
$\gamma(k, u_0)$, but this computation can be readily performed with
the aid of a symbolic algebra package such as {\it Mathematica}.  See
Appendix~\ref{app:serre} for an outline of the calculations for the
Serre system.  In fact, $\gamma$ is precisely the sought for nonlinear
frequency correction $\omega_2(k,\overline{u})$ mentioned earlier that
is obtained in a weakly nonlinear analysis of the Whitham equations.
Equations for $A_{T_1}$ and $A_{T_2}$ can be combined into a single
equation for the un-scaled envelope $\tilde A(x,t)$ defined by
\begin{equation}
\tilde A(x,t) = \eps A(\epsilon x,\epsilon t,\epsilon^2 t).
\end{equation}
Hence, one has the following substitution rules:
\begin{equation}
  \label{substitut}
  \tilde A_x = \eps^2 A_X ,\; \tilde A_t = \eps^2 A_{T_1} + \eps^3
  A_{T_2},\; |\tilde A|^2 = \eps^2 |A|^2 \, .
\end{equation}
The envelope of the wave packet $u=u_0+\tilde A(x,t)
e^{i(kx-\omega_0t)} + \rm{c.c.}$ is then governed by the equation:
\begin{equation}\label{NLS3}
i \tilde A_t + i  \partial_k\omega_0  \tilde A_x+ \beta \tilde A_{xx}
+\gamma |\tilde A|^2 \tilde A=0 \, .
\end{equation}
The sign of the product $\beta \gamma$ determines the NLS type: if
$\beta \gamma > 0$ the equation is attractive or focusing and
describes the envelope of a modulationally unstable wave while for
$\beta \gamma <0$ it is repulsive or defocusing and describes the
envelope of a modulationally stable wave.

To apply the NLS equation to the description of the DSW harmonic edge
vicinity, we assume in \eqref{exp1}:
$$
u_0 = u_-, \quad k=k_-, \quad \eps u_1= \tilde A(\chi, t) \exp[i(k_- x
- \omega_0(k_-,u_-)t)] + \hbox{c.c.},
$$ where $\chi=x-\partial_k\omega_0(k_-,u_-)\,t = x-s_-t$.
The DSW-NLS equation for $\tilde A$ is then
\begin{equation}\label{NLS2}
i \tilde A_t +\beta(k_-,u_-) \tilde A_{\chi\chi} +\gamma(k_-,u_-)
|\tilde A|^2 \tilde A=0,
\end{equation}
where the dependence of $k_-$ on the Riemann data $u_{\pm}$ is
obtained by DSW fitting.

We introduce the Madelung transformation $\tilde A= \sqrt{\rho}e^{i
  \phi}$, $v=\phi_\chi$ to cast the NLS equation (\ref{NLS2}) in
dispersive-hydrodynamic form
\begin{equation}\label{NLS_dh}
\begin{split}
&\rho_t+ 2 \beta (\rho v)_\chi = 0, \\
&v_t + 2\beta vv_\chi - \gamma \rho_\chi  -\beta  (\sqrt
\rho_{\chi\chi}/\sqrt{\rho})_\chi= 0,
\end{split}
\end{equation}
where  $v \sim k-k_-$, and $\sqrt \rho = |\tilde A| \sim a/4$ in the DSW
context (we recall that $a= u_{\rm max}- u_{\rm min}$). The DSW
modulation solution is a rarefaction curve of the Whitham equation so
the relevant NLS solution must also be a rarefaction wave described by
the long-wave, dispersionless limit.  Neglecting the dispersive term
in \eqref{NLS_dh} we obtain
\begin{equation}\label{sw}
\begin{split}
&\rho_t+ 2\beta (\rho v)_\chi = 0, \\
&v_t + 2 \beta vv_\chi - \gamma \rho_\chi  = 0.
\end{split}
\end{equation}
The characteristic velocities are $\beta v \pm \sqrt{-2\beta \gamma
  \rho}$, so the system is hyperbolic if $\beta \gamma<0$ and elliptic
if $\beta \gamma >0$, consistent with the defocusing and focusing
character of the NLS equation \eqref{NLS}, respectively.  We assume
for now that $\beta \gamma <0$, so that the system \eqref{sw} is
equivalent to the shallow water equations.

We now need to specify boundary conditions for the dispersionless NLS
equation \eqref{sw} at the DSW harmonic edge. This is done using the
GP matching conditions \eqref{gp_matching} and the DSW fitting
data. In the original modulation variables, we have from
\eqref{gp_matching}
\begin{equation} \label{bvp}
x = s_- t: \qquad a=0, \qquad k=k_-.
\end{equation}
Note that, unlike the free boundary in \eqref{gp_matching}, the
boundary in \eqref{bvp} is known from DSW fitting. Translating
\eqref{bvp} into the variables of \eqref{sw}, we arrive at a 
boundary value problem for the vacuum rarefaction wave
\begin{equation}
  \label{bvp1}
  \chi = 0, ~ t > 0: \quad \rho=0,  \quad v= 0.
\end{equation}

We are now looking for a self-similar rarefaction wave solution of the
shallow water equations \eqref{sw} subject to the boundary conditions
\eqref{bvp1}.  There are two such non-trivial solutions---the fast and
slow waves.  The correct one is chosen by the admissibility condition
that the wavenumber decrease monotonically as the DSW is traversed
from the harmonic to the soliton edge, i.e., $ dk/da < 0$ or,
equivalently, $ dv/d\rho < 0$.  Then the required solution of
\eqref{sw} is given by
\begin{equation}
  \label{NLS_sol}
  \sqrt{\rho} =\frac{1}{3 \sqrt{-2\beta \gamma}} \left|\frac{\chi}{t}\right|; \quad
  v= \frac{1}{3\beta} \frac{\chi}{t} \, .
\end{equation}
Using the dispersionless NLS solution \eqref{NLS_sol} and the
expansions \eqref{exp1}, we recover the leading order behaviors of the
amplitude $a $ and the wavenumber $k$ near the DSW harmonic edge in
terms of the dispersion and nonlinearity coefficients $\beta$ and
$\gamma$ of the associated NLS equation \eqref{NLS2}
\begin{equation}\label{mod_general_harmonic1}
a \sim  \frac{4}{3}
\frac{1}{\sqrt{-2\beta \gamma}}|x/t - s_-| ,\quad
k -k_-\sim \frac{1}{3\beta} \left( x/t - s_-\right).
\end{equation}
We also recover the variation of the mean:
\begin{equation}\label{ubar_NLS}
  \overline{u} - u_-\sim b_1(k_-,u_-) \, \rho \propto \left(x/t -
    s_-\right)^2. 
\end{equation}
Equation~\eqref{mod_general_harmonic1} is the universal description of
the DSW envelope (with ``martini-glass'' shape
\cite{kodama_whitham_2009,el_dispersive_2016}) near the harmonic edge
for systems with convex dispersion ($\beta \ne 0$).

 Solution \eqref{mod_general_harmonic1} is valid when $\beta \gamma <0$, which 
is the hyperbolicity condition for the dispersionless NLS system \eqref{sw} and can be interpreted as a necessary condition for DSW modulational stability. 
 For non-convex dispersion relations, the
sign of $\beta$ can change and the system may exhibit an unstable
behavior described by the focusing NLS equation where $\beta \gamma
>0$. An example of such behavior has been recently reported in
\cite{maiden_modulations_2016}, where it was shown that nonconvexity
of the linear dispersion relation for the conduit equation implies an
elliptic regime for the associated Whitham equations in a certain
region of parameter phase space.  This gives rise to modulationally
unstable dynamics that can be described by the focusing NLS equation
in the small amplitude regime.

For systems with non-convex dispersion, the study of DSW behavior near
the zero dispersion point $\beta=0$ necessitates inclusion of higher
order terms in the associated NLS equation.  It turns out that the
inclusion of such terms is beneficial even outside the non-convex,
zero-dispersion regime, as we now demonstrate.

\section{Higher order NLS approximation}
\label{sec:higher_order_nls}

We can include higher order nonlinear/dispersive effects in the
asymptotic expansion \eqref{exp1} by introducing a third, slower time
scale $T_3=\eps^3 t$ and assuming $u_1 =
A(X,T_1,T_2,T_3)e^{i(kx-\omega t)}+ {\rm c.c.}$ The cancellation of
secular terms at $O(\eps^4)$ then gives
$$
A_{T_3} + \delta A_{XXX} + \lambda |\tilde A|^2 A_X + \mu A^2 A^*_X = 0\, ,
$$
where $\delta(k,u_0)=-\partial_{kkk}\omega_0/6$ and $A^*$ is the
complex conjugate of $A$. Once again, the laborious part of the
computation consists in finding the coefficients
$\lambda(k,u_0)$ and $\mu(k,u_0)$.
% {\color{red}(see Appendix~\ref{app:conduit} for the relevant
% computations for the conduit equation performed using {\it
% Mathematica})}
Written in the moving reference frame $(\chi,
t)$ (cf.~Appendix~\ref{app:HNLS_conduit} for the derivation of the
un-scaled equation), where $\chi=x - \partial_k\omega_0(k, u_0) t$
(recall Eq.~\eqref{NLS2}), the resulting equation is the higher order NLS, or HNLS,
equation
\begin{equation}\label{HNLS}
i \tilde A_t +  \beta  \tilde A_{\chi\chi} + \gamma | \tilde A|^2
\tilde A  + i \delta
  \tilde A_{\chi\chi\chi} + i \lambda  |\tilde A|^2
 \tilde A_\chi+ i \mu  \tilde A^2  \tilde A^*_\chi = 0,
\end{equation}
initially derived as an improvement to the standard NLS equation for
signal propagation in optical fibers \cite{kodama_nonlinear_1987}; 
it also arises in geophysical fluid dynamics
\cite{grimshaw_long-time_2008} and other areas.  See
\cite{sedletsky_fourth-order_2003} and references therein. In the
context of the description of a DSW, we set $u_0=u_-$ and
$k=k_-(u_-, u_+)$ from DSW fitting in the coefficients $\beta$,
$\gamma$, $\delta$, $\lambda$, $\mu$.

Similar to the previous section, we cast the HNLS
equation~\eqref{HNLS} in dispersive hydrodynamic form using the
Madelung transform $\tilde A = \sqrt \rho e^{i\int v \,d\chi}$.  Upon
neglecting dispersive terms, we obtain the following quasilinear
system for long waves
\begin{equation}
  \label{HNLS_displess}
  \begin{split}
    &\rho_t + \left(2\beta \rho v-3\delta \rho v^2 +(\lambda+\mu)\rho^2/2
    \right)_\chi = 0,\\
    &v_t + \left( \beta v^2 - \delta v^3 - \gamma \rho +(\lambda-\mu) \rho v
    \right)_\chi  =0.
  \end{split}
\end{equation}
As expected, the system \eqref{HNLS_displess} is equivalent to the
shallow water equations \eqref{sw} when $\delta=\lambda=\mu=0$. In the
context of DSWs, the dispersionless limit \eqref{HNLS_displess} of the
HNLS equation should be considered with the same vacuum rarefaction
conditions \eqref{bvp1} augmented by the DSW admissibility inequality $dv/d\rho<0$.

Before we proceed with the integration of system
\eqref{HNLS_displess}, let us briefly discuss its structure.  The
characteristic velocities are
\begin{equation}
  \label{Vsigma}
  \begin{split}
    V_\pm (\rho,v) &= \lambda \rho + 2\beta v -3\delta v^2 \pm \rho
    \sqrt{D(\rho, v)}, \\
    D(\rho,v) &= \mu^2 - 2 (\beta - 3 \delta v)(\gamma-(\lambda-\mu)v)/\rho
  \end{split} .
\end{equation}
and the associated right eigenvectors are
\begin{equation}
  \mathbf{R}_\pm = \left(-\mu \rho \pm \rho \sqrt{D(\rho,v)},\gamma-(\lambda-\mu)v
  \right)^T,
\end{equation}
implying that the system \eqref{HNLS_displess} is hyperbolic in the
region $(\rho,v)$ where $\rho > 0$, the discriminant is positive $D >
0$, and $\gamma \ne (\lambda-\mu)v$ so that $\mathbf{R}_\pm$ are independent. In the small amplitude regime ($\rho,|v| \ll 1$), we
recover the standard hyperbolicity condition $\beta \gamma < 0$ since
$\beta, \gamma \ne
0$.  % genuinely nonlinear \cite{lax_hyperbolic_1973} ($\nabla V_\pm
% \cdot R_\pm \neq 0$) if
% \begin{equation}
%   \begin{split}
%     &D
%     [\mu(\lambda+\mu)\rho -3(\beta-3\delta v)(\gamma-(\lambda-\mu)v)]^2\\
%     &+[\mu^2(\lambda+\mu)\rho-3 (\gamma\delta+ \beta\lambda-
%     \delta(4\lambda-\mu )v)(\gamma-(\lambda-\mu)v)]^2\neq 0\,.
%   \end{split}
% \end{equation}
%\GE{Consider removing} Assuming hyperbolicity, verified {\it a posteriori}, the system \eqref{HNLS_displess} is genuinely
%nonlinear \cite{lax_hyperbolic_1973} ($\nabla V_\pm \cdot R_\pm \neq
%0$) if and only if
%\begin{equation}
%  \label{eq:1}
%  \begin{split}
%    \frac{3}{2} D + \mu \lambda - \frac{1}{2} \mu^2 &\ne 0 , \quad
%    (\lambda-\mu)(4 \delta v - \beta) - \delta \gamma \ne 0 .
%    % \mu(\lambda+\mu)\rho -3(\beta-3\delta v)(\gamma-(\lambda-\mu)v)
%    % &\ne 0 \\
%    % \mu^2(\lambda+\mu)\rho-3 (\gamma\delta+ \beta\lambda-
%    % \delta(4\lambda-\mu )v)(\gamma-(\lambda-\mu)v) &\ne 0 
%  \end{split}
%\end{equation}
Consequently, the DSW modulation near the harmonic edge is determined
by the similarity solution of \eqref{HNLS_displess},
\begin{equation}
  \label{Vd_sim}
  V_d(\rho, v(\rho))=\frac{\chi}{t},
\end{equation}
where $d= -\hbox{sgn} \ \beta$ is the DSW orientation and the
dependence $v(\rho)$ is determined by the characteristic ODE
\begin{equation}\label{ODE_HNLS}
\frac{d v}{d \rho} +
\frac{\mu - d \sqrt{ D(\rho,v)}}{2(\beta - 3 \delta v)} = 0 , \quad v(0)=0.
\end{equation}

As a by-product of the multi-scale expansion to order $O(\eps^4)$, we
obtain a higher order correction to the mean value (recall
\eqref{ubar_NLS}) described by the new expression
\begin{equation}\label{ubar2}
  \ubar  - u_-  \sim  \rho(b_1 + b_2 v),
\end{equation}
where $b_1$ has been obtained at the previous order ($O(\eps^3)$), and
$b_2$ is determined as a by-product of the $O(\eps^4)$ computation.

We now obtain the second-order expansion of the simple wave modulation
solution \eqref{Vd_sim}, \eqref{ODE_HNLS} near $\chi=0$ for small
$\rho$, $v$, which will improve the first-order NLS result
\eqref{mod_general_harmonic1} in the vicinity of the harmonic edge.
Seeking the solution of \eqref{Vd_sim}, \eqref{ODE_HNLS} in the form
of a series in powers of $\chi/t$, we obtain universal asymptotic
expressions for the DSW amplitude and wavenumber modulations
(cf.~\eqref{mod_general_harmonic1})
\begin{equation}
\label{dexplicit}
\begin{split}
& a =  \frac{4}{3 \sqrt{-2 \beta
\gamma}} \left|\frac {\chi} t \right| + \hbox{sgn} \ \beta \ \frac{4(2\gamma
\delta / \beta + \lambda -\mu)}{9 (-2 \beta
\gamma)^{3/2}} \left(\frac \chi t \right)^2 + O\left[\left(\frac \chi
t \right)^3 \right],\\
&k = k_- + \frac{1}{3\beta} \frac \chi t + \frac{7\delta +\beta\lambda
/\gamma}{36 \beta^3} \left(\frac \chi t \right)^2 + O\left[\left(\frac \chi
t \right)^3 \right].
\end{split}
\end{equation}
We note that it is implicit in the expansions \eqref{dexplicit} that
$|\beta(k_-, u_-)|=O(1)$, which is the case for dispersive
hydrodynamic equations with a convex dispersion relation. For systems
with non-convex dispersion such as the Benjamin-Bona-Mahony equation \cite{benjamin_model_1972} or the conduit equation \cite{scott_observations_1986, lowman_dispersive_2013}, the behaviour near the zero-dispersion point, $\beta(k_-, u_-)=0$ captured by HNLS equation \eqref{HNLS} requires a separate consideration, which is beyond the scope of the present paper.

%an expansion near the zero-dispersion
%point, $\beta(k_-, u_-)=0$, leads to a different (quadratic in
%$\chi/t$ \GE{Check!}) leading order DSW envelope behavior in the
%vicinity of the harmonic edge. Importantly, the full simple wave
%solution \eqref{Vd_sim}, \eqref{ODE_HNLS} is valid in both of the
%convex and non-convex cases so long as \eqref{HNLS_displess} remains
%hyperbolic and genuinely nonlinear.

For convex dispersive hydrodynamics, the second order approximation
\eqref{dexplicit} formally delivers the same accuracy in the
description of the vicinity of the DSW harmonic edge as the HNLS
equation \eqref{HNLS} itself.  However, comparisons with results of
direct numerical simulations of the Riemann problem for the example
dispersive hydrodynamic equations in the next Section show that the
simple wave solution \eqref{Vd_sim}, \eqref{ODE_HNLS} of the full
dispersionless HNLS \eqref{HNLS_displess}  exhibits better
agreement with numerical solution than the expansion
\eqref{dexplicit}. Remarkably, this agreement holds over a significant
portion of a DSW, where the amplitude is not small and the HNLS
description, let alone the expansions \eqref{dexplicit}, are formally
not expected to be applicable.

\section{NLS description of dispersive shock waves:
Examples}
\label{sec:nls_examples}

We now demonstrate the effectiveness of the developed general approach by
applying it to several specific dispersive hydrodynamic systems and
comparing the results with direct numerical simulations of the
corresponding Riemann problems.

\subsection{Korteweg-de Vries equation}
\label{sec:kdv}

As a first example, we consider the KdV equation
\begin{equation}\label{kdv}
u_t + u u_x + u_{xxx} = 0 
\end{equation}
with Riemann initial data \eqref{eq:riemann_data1}. The aim here is to
compare the results of the developed asymptotic approach with the
known GP modulation solution \cite{gurevich_nonstationary_1974}.

The KdV linear dispersion relation is
\begin{equation}
  \label{disp_kdv}
  \omega_0(k,u_0) = u_0 k -k^3 .
\end{equation}
The multiple scales asymptotic expansion of~\eqref{kdv} leading to the
NLS equation for KdV weakly nonlinear wavepackets is standard and can
be found in the literature, see e.g.~\cite{ablowitz_nonlinear_2011,
  boyd_weakly_2001}. The coefficients in~\eqref{NLS}
and~\eqref{ubar_NLS} are
\begin{equation}\label{coef_kdv}
\beta = -3 k, \quad\gamma = \frac{1}{6k},\quad b_1 = -\frac{1}{3k^2}.
\end{equation}
We also derive the coefficients of the HNLS equation~\eqref{HNLS} and
the associated higher order correction of the mean
value~\eqref{ubar2} \cite{boyd_weakly_2001}
\begin{equation}\label{bbm_nls}
\delta = 1 ,\quad \lambda= -\frac{1}{3k^2}, \quad \mu = - \frac{1}{2k^2}, \quad
b_2=\frac{2}{3k^3}.
\end{equation}
The trailing edge wavenumber is readily obtained from DSW fitting (see
\cite{el_resolution_2005} and Sec.~\ref{sec:dsw_fitting}). Solving the
ODE \eqref{ode_char}, we obtain $k_-=K(u_-)$ in the form
\begin{equation}\label{bbm_kmin}
  k_ - = \sqrt{\frac23 \Delta}  \quad\text{with}\quad \Delta = u_- - u_+
  \, ,
\end{equation}
which yields the harmonic edge velocity 
\begin{equation}
  \label{eq:2}
  s_- = \partial_k \omega_0(k_-,u_-) = u_- - 2 \Delta \, .
\end{equation}
Substituting \eqref{bbm_nls}, \eqref{bbm_kmin} into \eqref{ubar2} ,
\eqref{dexplicit}, we obtain the second order expansions
\begin{eqnarray}
a &=& \frac43 \frac\chi t - \frac{1}{27 \Delta} \left( \frac\chi t
\right)^2 + O \left[\left( \frac\chi t \right)^3
\right], \label{a_kdv}\\
k &=& \sqrt{\frac{2\Delta}{3}} - \frac{1}{3\sqrt{6\Delta}} \frac \chi
t - \frac{13}{216 \sqrt{6} \Delta^{3/2}} \left(\frac \chi t \right)^2
+O\left[ \left(\frac \chi t \right)^3\right], \label{k_kdv} \\
\ubar &=& u_- - \frac{1}{18 \Delta} \left(\frac \chi t \right)^2 -
\frac{5}{324 \Delta^2} \left(\frac \chi t \right)^3 +O\left[
\left(\frac \chi t \right)^4 \right], \label{ubar_kdv}
\end{eqnarray}
which agree with the corresponding expansions of the exact GP solution
\cite{gurevich_nonstationary_1974}. 
\begin{figure}
%\begin{subfigure}[t]{.33\textwidth}
%\centering
\includegraphics[width=0.5\textwidth]{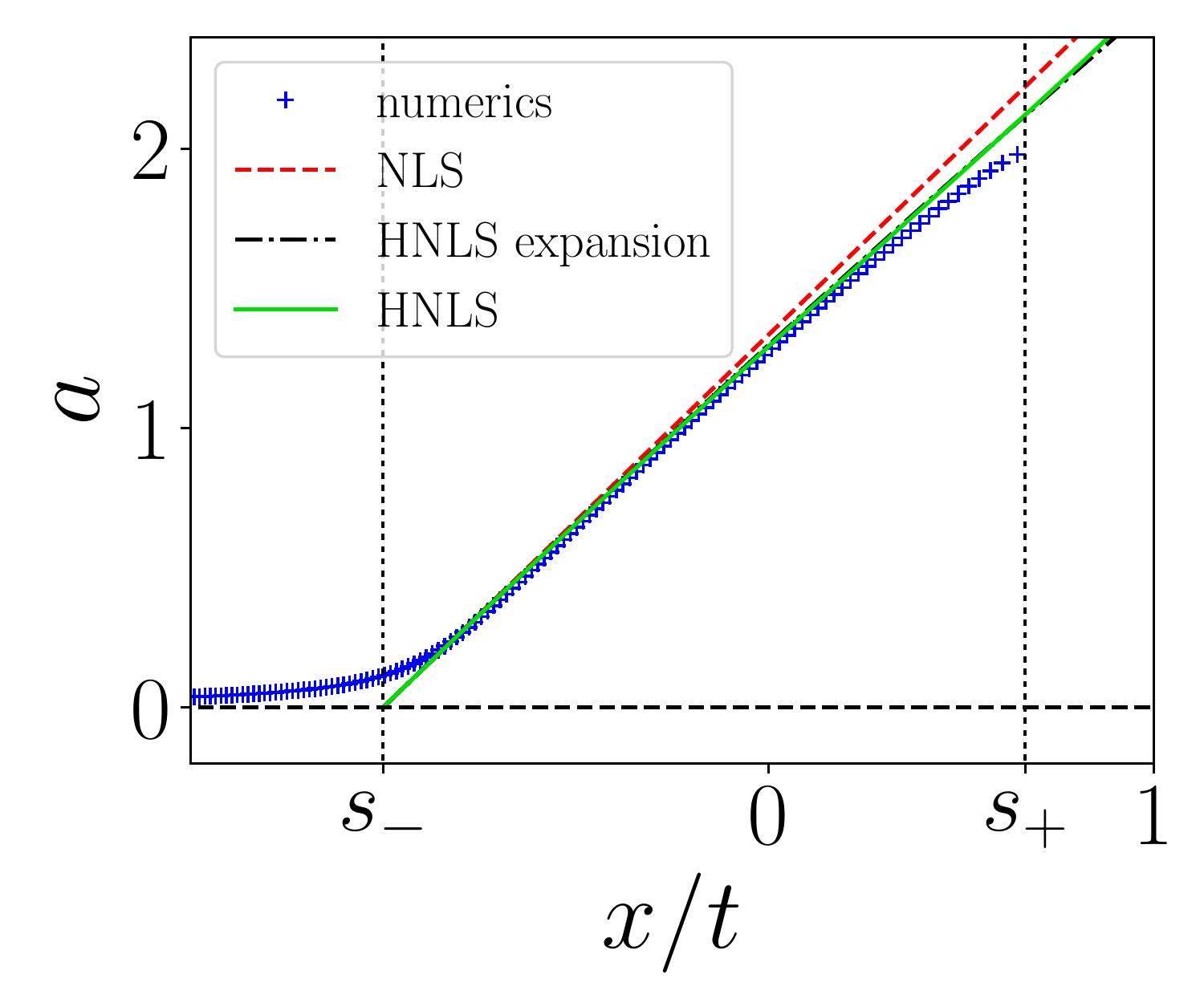}
%\end{subfigure}%
%\begin{subfigure}[t]{.33\textwidth}
%\centering
\includegraphics[width=0.5\textwidth]{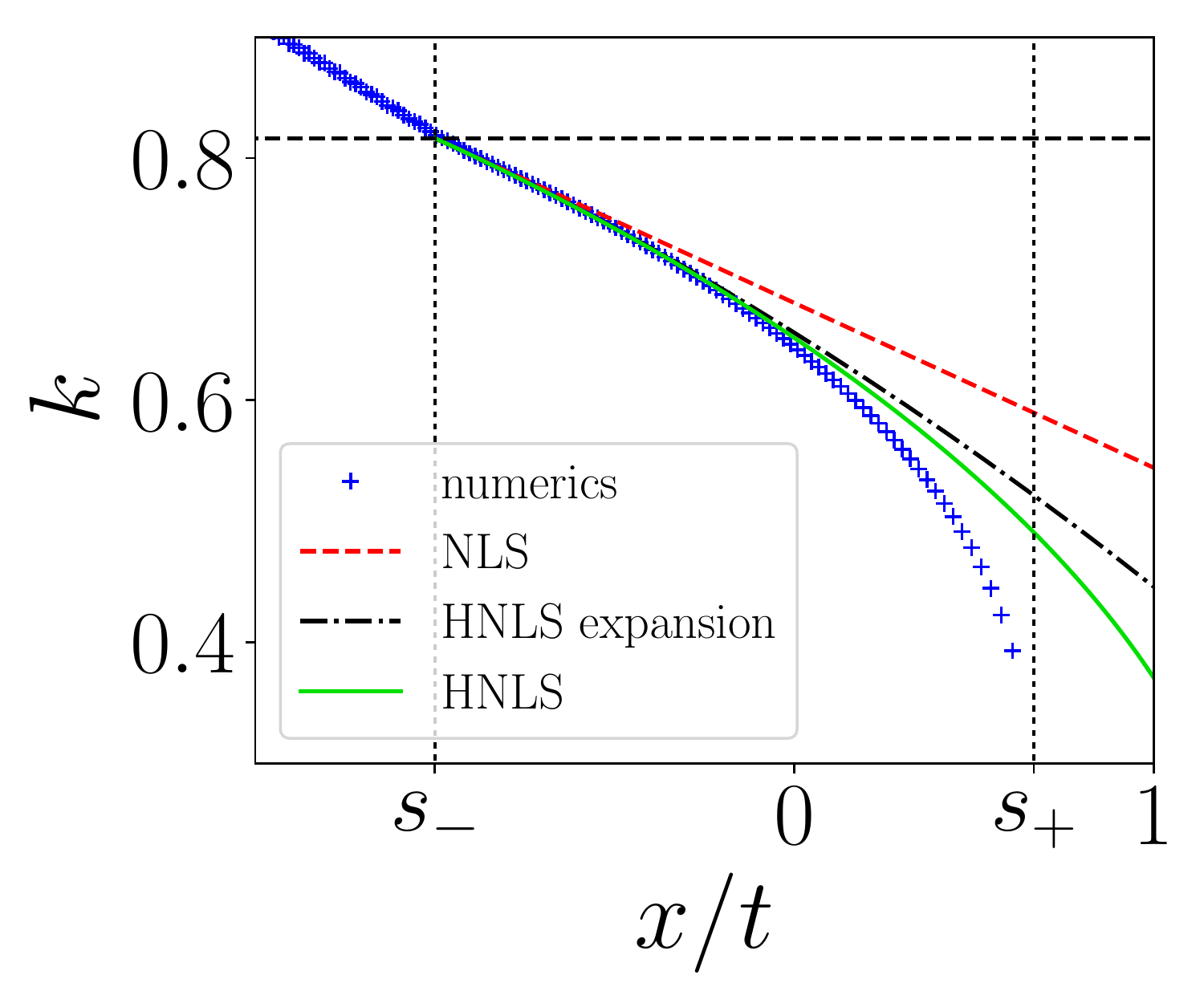}
%\end{subfigure}
%\begin{subfigure}[t]{.33\textwidth}
%\centering
\includegraphics[width=0.5\textwidth]{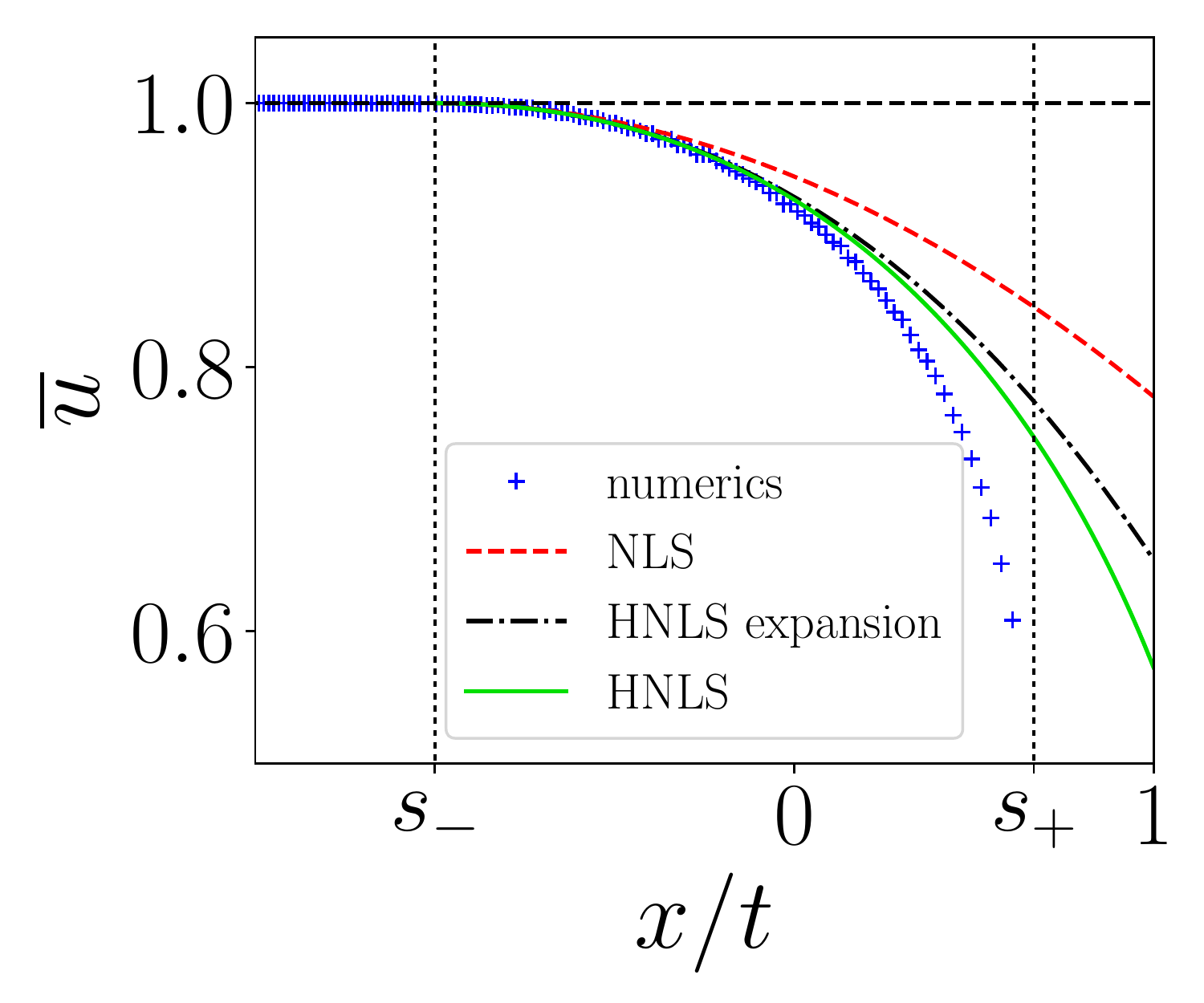}
%\end{subfigure}
\caption{Amplitude, wavenumber, and mean field profiles for the KdV
  Riemann problem with $u_-=1$ and $u_+=0$. Comparison between direct
  numerical simulations at $t=500$ (blue pluses), the NLS/HNLS
  asymptotic descriptions given by (i) expansions
  \eqref{a_kdv}--\eqref{ubar_kdv} with the first order (NLS)
  approximation (dashed red line) and the second order (HNLS)
  approximation (dash-dotted black line) and (ii) the exact simple wave
  solution Eqs.~\eqref{Vd_sim}, \eqref{ODE_HNLS} of the dispersionless
  HNLS equation (solid green line). The horizontal black dashed lines
  correspond to the values of the corresponding fields in the harmonic
  limit: $a=0$, $k=k_-$ and $\ubar = u_-$.}
\label{fig:kdv}
\end{figure}

The comparison between the exact simple wave solution \eqref{Vd_sim},
\eqref{ODE_HNLS} of the dispersionless HNLS equation
\eqref{HNLS_displess} with coefficients \eqref{coef_kdv}, the
asymptotic expansion \eqref{a_kdv}--\eqref{ubar_kdv}, and the direct
numerical solution of the KdV Riemann problem is displayed in
Fig.~\ref{fig:kdv}.

The numerical method used in the simulations is detailed in
Appendix~\ref{app:num}. Although the NLS description is formally
limited to the small-amplitude regime in the vicinity of the DSW
harmonic edge, the KdV DSW amplitude is almost linear for $x/t \in
[s_-,s_+]$, so the term linear in $x/t$ in~\eqref{a_kdv} fits almost
the entire DSW with good accuracy. It is no longer the case for the
wavenumber and the mean flow: good agreement between numerics and the
first order (NLS) approximation is observed only in the vicinity of
the trailing edge, but the second order expansions \eqref{k_kdv},
\eqref{ubar_kdv} exhibit better agreement with the direct KdV numerics
over a broader $x/t$ interval. The agreement further improves when the
full solution~\eqref{Vd_sim}, \eqref{ODE_HNLS} of the dispersionless
HNLS equation is used where all modulation parameters fit almost the
entire DSW with good accuracy, with the exception of some vicinity of the leading edge, where $k \to 0$
and $\ubar \to u_+$ logarithmically in $\chi/t$ \cite{gurevich_nonstationary_1974},  behavior that cannot be captured by the expansions \eqref{k_kdv}, \eqref{ubar_kdv}.

We should make an important comment regarding the comparison of the
behavior of the DSW oscillations near the harmonic edge with the
results of modulation theory. We can see from Fig.~\ref{fig:kdv}, left
panel, that there is some deviation of the envelope profile in a close
vicinity of the trailing edge from linear behavior $a \propto
(x/t-s_-)$ that is predicted by modulation theory. In particular, the
amplitude of the oscillations is not exactly zero for $x/t \le s_-$.
This discrepancy between the DSW modulation solution and the exact
oscillation behavior is known, having been studied for the KdV
equation in detail in \cite{grava_numerical_2007} where it was shown
that the envelope amplitude difference between the numerical KdV
solution and the modulation theory solution in the region $x/t < s_-$
decreases roughly as $t^{-1/3}$.  However, typically, modulation
theory provides a very satisfactory prediction of the amplitude growth
near the DSW harmonic edge even for relatively moderate times.

%{\color{red} \it Remark: the solid line representations of the
%amplitude and of the wavenumber in Fig.~\ref{fig:kdv} (and
%Figs.~\ref{fig:serre}, \ref{fig:GNLS}, \ref{fig:BBM} in the following)
%correspond respectively to the implicit plots $(f(\rho), \sqrt \rho)$ and
%$(g(v), k_0+v)$ [cf. Eq.~\eqref{def_fg}]. It appears that the
%implicit representation of $\rho(\chi/t)$ and $v(\chi/t)$ fits the numerics
%much better than the explicit representation obtained by
%solving~\eqref{rho_HNLS} and~\eqref{v_HNLS}:
%\begin{equation}
%\label{dexplicit}
%\begin{split}
%&\sqrt \rho \simeq  - \frac{\sigma}{3 \sqrt{-2 \beta
%\gamma}} \frac \chi t - \sigma \frac{2\gamma
%\delta / \beta + \lambda -\mu}{9 (-2 \beta
%\gamma)^{3/2}} \left(\frac \chi t \right)^2 + O\left[\left(\frac \chi
%t \right)^3 \right],\\
%&v \simeq \frac{1}{3\beta} \frac \chi t + \frac{7\delta +\beta\lambda
%/\gamma}{36 \beta^3} \left(\frac \chi t \right)^2 + O\left[\left(\frac \chi
%t \right)^3 \right].
%\end{split}
%\end{equation}
%Indeed the description provided by the HNLS equation happens to fit
%the numerics even outside the domain of validity of the asymptotic
%method where: $\rho, v \sim 1$. In that case the self-similar
%parameter $\chi /t$ given
%by~\eqref{rho_HNLS} [or~\eqref{v_HNLS}] can no longer be
%considered as small and~\eqref{dexplicit} fails to describe the DSW far from
%the harmonic edge.
%}

\subsection{Conduit equation}
\label{sec:conduit}

We now consider the conduit equation
\begin{equation}
  \label{conduit}
  u_t+2uu_x-\left[ u^2 \left(u_t/u \right)_x \right]_x = 0\, ,
\end{equation}
a non-integrable example that gives rise to unstable behavior not
found in the KdV equation.  This equation approximately models the
evolution of the cylindrical interface, with cross-sectional area $u$ at
time $t$ and vertical spatial coordinate $x$, separating a light,
viscous fluid rising buoyantly through a heavier, more viscous,
miscible fluid at small Reynolds numbers
\cite{scott_observations_1986, lowman_dispersive_2013}.

Equation \eqref{conduit} has convex hyperbolic flux $f(u) = u^2$ and
linear dispersion relation
\begin{equation}
\omega=\omega_0(k,u_0) = \frac{2 u_0 k}{1+u_0k^2},
\end{equation}
which is non-convex as $\partial_{kk} \omega_0$ can change sign.
%The group velocity and the dispersion relation curvature are
%\begin{equation}
%\partial_k \omega_0 = \frac{2u_0(1-u_0 k^2)}{(1+ u_0 k^2)^2}, \qquad
%\partial_{kk} \omega_0 = \frac{ 4u_0^2 k (u_0 k^2-3)}{(1+ u_0 k^2)^3}.
%\end{equation}
%One can see that $\partial_{kk} \omega_0=0$ when $k=0$ or
%$k=\sqrt{3/u_0}$, or $u_0 =0$, so the dispersion relation is
%non-convex, unlike the dispersion relation for the KdV equation
%\eqref{disp_kdv} and the Serre equations \eqref{ldr_serre}. One of the main
%implications of non-convex dispersion of the conduit equation is the
%effect of the DSW implosion observed in \cite{lowman_dispersive_2013}
%for a certain region of initial data parameters. Here we shall use the
%developed (H)NLS DSW analysis to explain and quantify this
%observation.

\begin{figure}

%\begin{subfigure}[t]{.33\textwidth}
%\centering
%\includegraphics[width=\textwidth]{fig/conduit_limit}
%\end{subfigure}
%\begin{subfigure}[t]{.33\textwidth}
\includegraphics[width=0.45\textwidth]{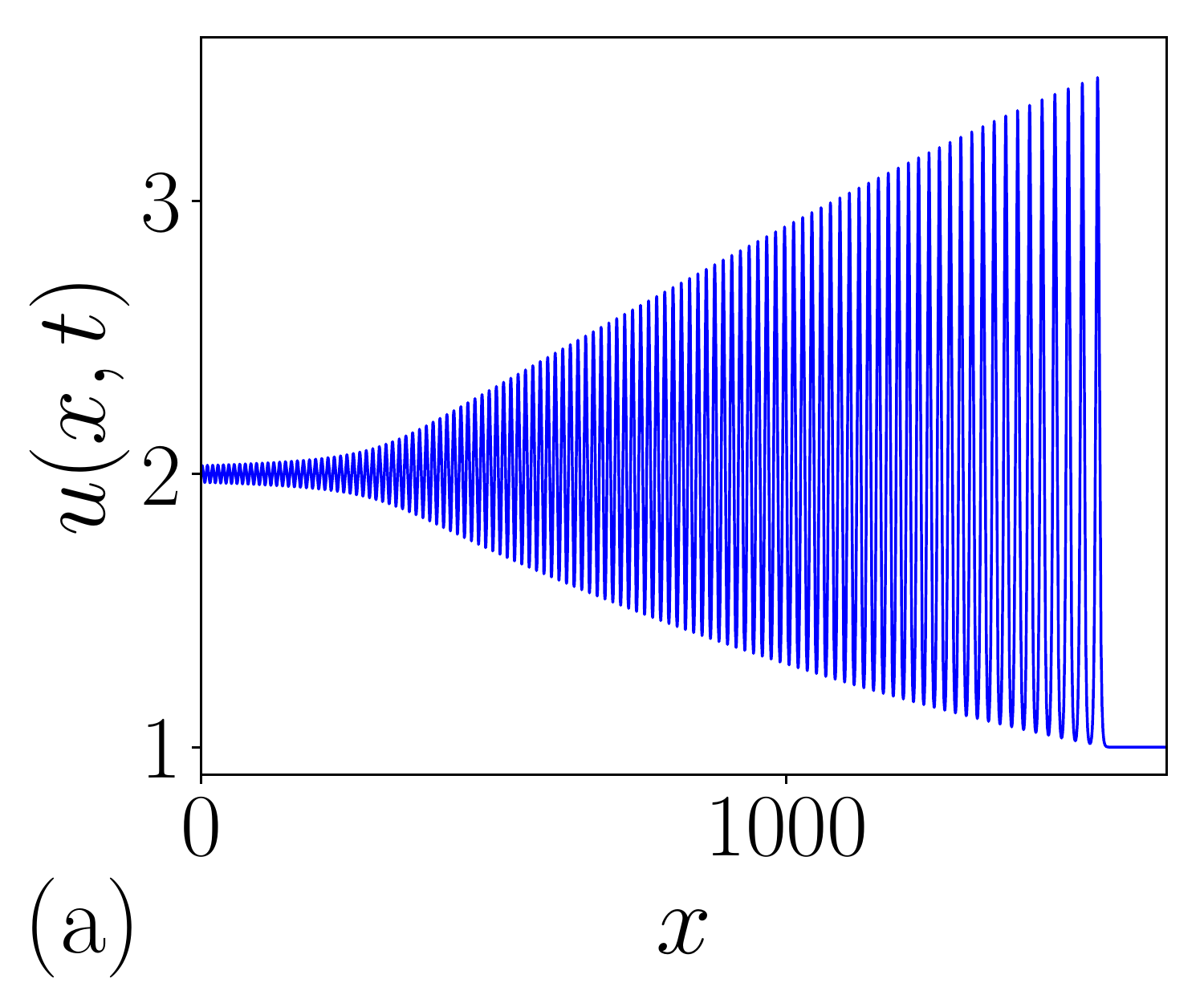}
%\end{subfigure}%
\quad
%\begin{subfigure}[t]{.33\textwidth}
%\centering
\includegraphics[width=0.45\textwidth]{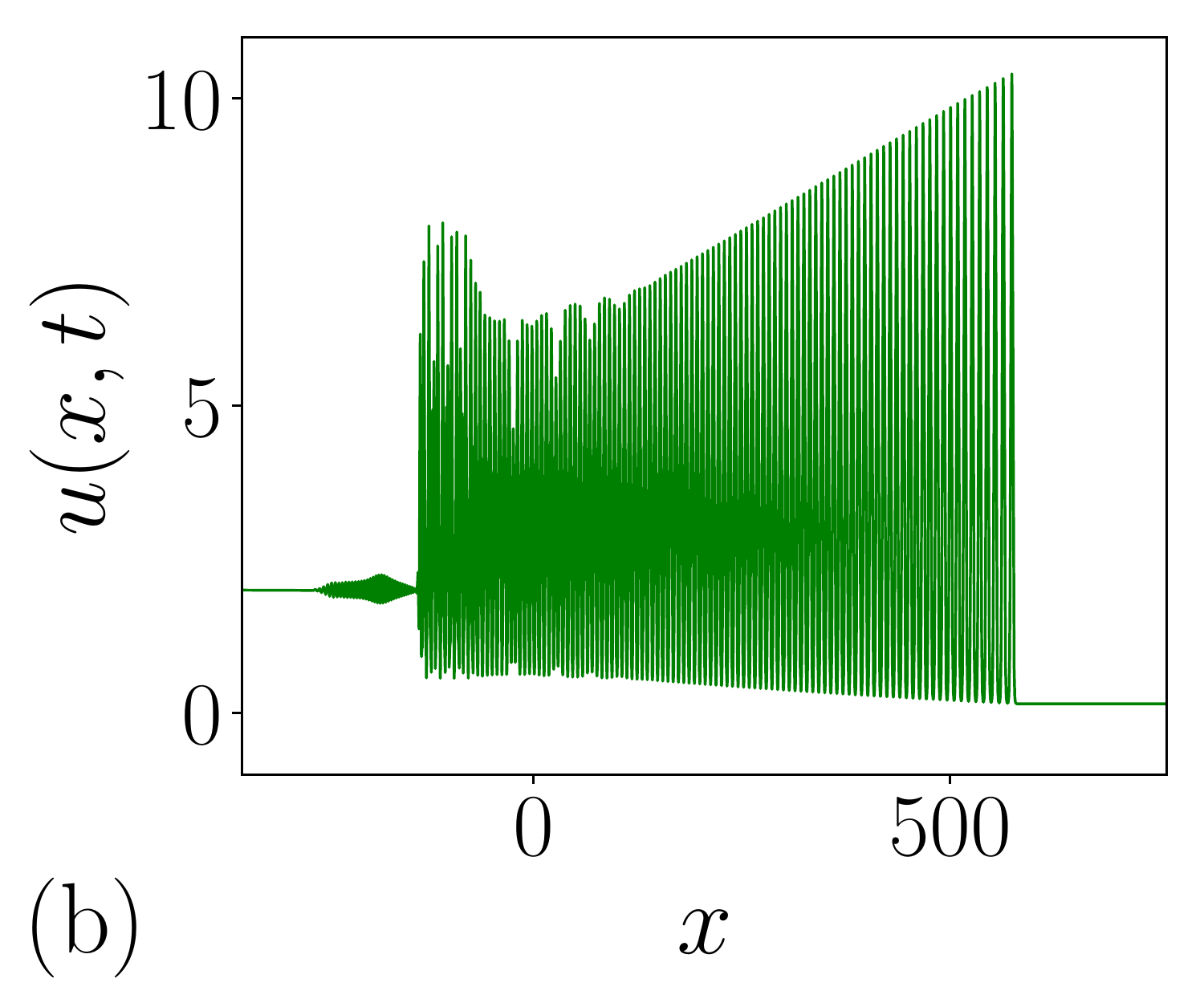}
%\end{subfigure}
\caption{Conduit DSW stability determined by the value of $u_+/u_-$.
 a)  Stable DSW with $u_+=1, u_-=2$ so that $u_+/u_-=1/2>
  5/32$;  b) Unstable, imploded DSW with $u_+=0.15, u_-=2$ so that
  $u_+/u_-<5/32$.}
\label{fig:conduit}
\end{figure}

The coefficients of the NLS equation Eq.~\eqref{NLS} and the
associated mean for Stokes waves of the conduit equation were derived
in Ref.~\cite{maiden_modulations_2016} and are
\begin{equation}\label{conduit_coef}
  \begin{split}
    &\beta = -\frac{2u_0^2k(3-u_0k^2)}{(1+u_0k^2)^3},\\
    &\gamma = \frac{3+5u_0k^2+8u_0^2k^4}{u_0^2k(9+12u_0k^2+3u_0^2k^4)},\\
    &b_1=-\frac{(1+u_0k^2)(1-3u_0k^2)}{u_0^2k^2(3+u_0k^2)}.
  \end{split}
\end{equation}
We see that, while the nonlinearity coefficient $\gamma$ is always
positive, the dispersion coefficient $\beta$ (and therefore the
parameter $\beta \gamma$) can change sign, so the parameter space
$(k, u_0)$ of conduit Stokes waves is split into two domains---which
correspond to the defocusing and focusing NLS regimes---that are
separated by the line $k=\sqrt{3/u_0}$ . In the context of DSWs, the
line $k_-(u_-, u_+)=\sqrt{3/u_-}$ in the $u_-$-$u_+$ phase plane of
Riemann data \eqref{eq:riemann_data1} separates the regimes of DSW
stability and instability. Here,
\begin{equation}
  \label{conduit_k0}
  k_- = \frac12 \sqrt{\frac{1}{u_+}-\frac{4}{u_-}+\sqrt{\frac{1}{u_+}
      \left(\frac{1}{u_+} + \frac{8}{u_-} \right) }} 
  % k_- = \frac12 \sqrt{\frac{1}{u_+}-4+\sqrt{\frac{1}{u_+}
  % \left(\frac{1}{u_+} + 8 \right) }} .
\end{equation}
is the conduit DSW harmonic edge wavenumber obtained from DSW fitting
\cite{lowman_dispersive_2013}.  The condition $k_- <\sqrt{3/u_-}$ or,
equivalently, $u_+/u_- > 5/32 \approx 0.156$ is the DSW fitting
admissibility condition whose violation was associated in
\cite{lowman_dispersive_2013} with a gradient catastrophe for the
wavenumber and a subsequent DSW implosion---the formation of a
two-phase region near the trailing edge.  Within the NLS description
of DSW modulations developed here, the above admissibility condition
is naturally interpreted as conduit DSW modulational stability
condition.  The plots of DSWs for stable and unstable regimes are
presented in Fig.~\ref{fig:conduit} (a) and (b) respectively.
%We note that the harmonic
%edge wavenumber $k_-$ is not defined in the unstable regime so the
%formally determined edge speeds $s_{\pm}$  do not coincide with the actual boundaries
%of the imploded DSW.

We now compare the predictions of the (H)NLS analysis for the conduit
DSW modulation with direct numerical simulations within the admissible
range of Riemann data with $u_+/u_- > 5/32$ that produces a stable
DSW. Within this region, $\beta<0$ so the DSW orientation $d= +1$, and
the harmonic edge is the trailing one, see Fig.~\ref{fig:conduit}.
The coefficients $\delta$, $\lambda$, $\mu$ in the conduit-HNLS
equation \eqref{HNLS} and the coefficient $b_{2}$ in the second order
expansions of the mean flow $\ubar$ are derived using symbolic
computations in {\it Mathematica} as presented in Appendix
\ref{app:HNLS_conduit} (see formulae \eqref{conduit_coef2}).

%Equation ~\eqref{conduit} has all the necessary pre-requisites for the
%application of the DSW fitting with some restrictions on admissible
%values of the Riemann data \eqref{eq:riemann_data1}.  Below we take advantage of the
%results of \cite{lowman_dispersive_2013}, where the DSW fitting procedure for the conduit
%equation has been realised.
%
%We consider the Riemann data with $0< u_+ < u_-$. We first assume $\partial_{kk} \omega_0(k_-, u_-)<0$ for which the DSW has the
%orientation $d =
%+1$. 
%The wavenumber of the trailing edge $k_-$ derived in
%~\cite{lowman_dispersive_2013} is given by:
%\begin{equation}\label{conduit_k0}
%k_- = \frac12 \sqrt{\frac{1}{u_+}-\frac{4}{u_-}+\sqrt{\frac{1}{u_+}
%\left(\frac{1}{u_+} + \frac{8}{u_-} \right) }} .
%% k_- = \frac12 \sqrt{\frac{1}{u_+}-4+\sqrt{\frac{1}{u_+}
%% \left(\frac{1}{u_+} + 8 \right) }} .
%\end{equation}
% One can also derive the coefficients of the HNLS
% equation~\eqref{HNLS} and the higher order correction of
% $(\overline{\eta},\ubar)$~\eqref{ubar2} following the procedure detailed in
% Appendix~\ref{app:BBM} for the BBM equation:
\begin{figure}
%\begin{subfigure}[t]{.33\textwidth}
%\centering
\includegraphics[width=0.5\textwidth]{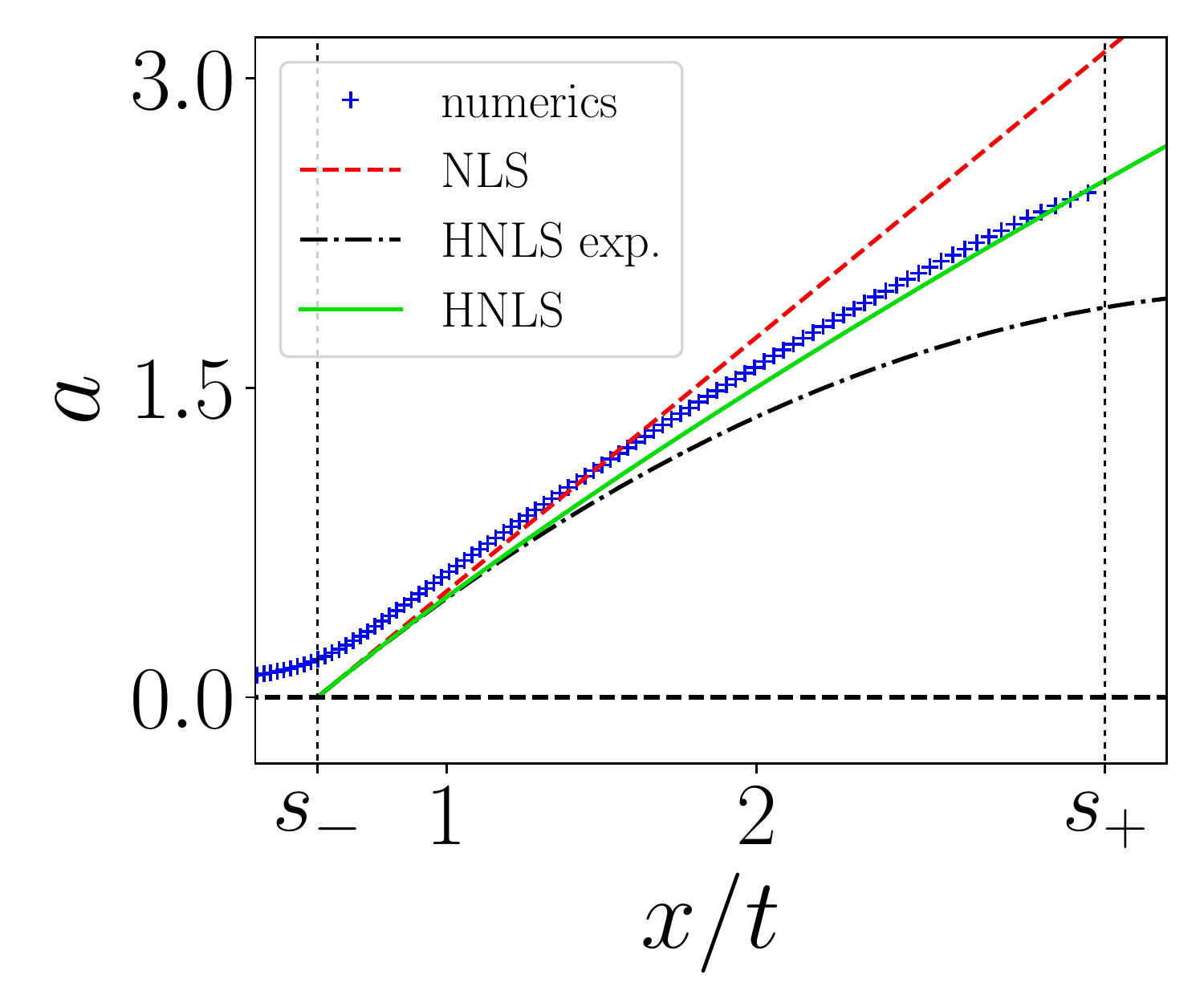}
%\end{subfigure}%
%\begin{subfigure}[t]{.33\textwidth}
%\centering
\includegraphics[width=0.5\textwidth]{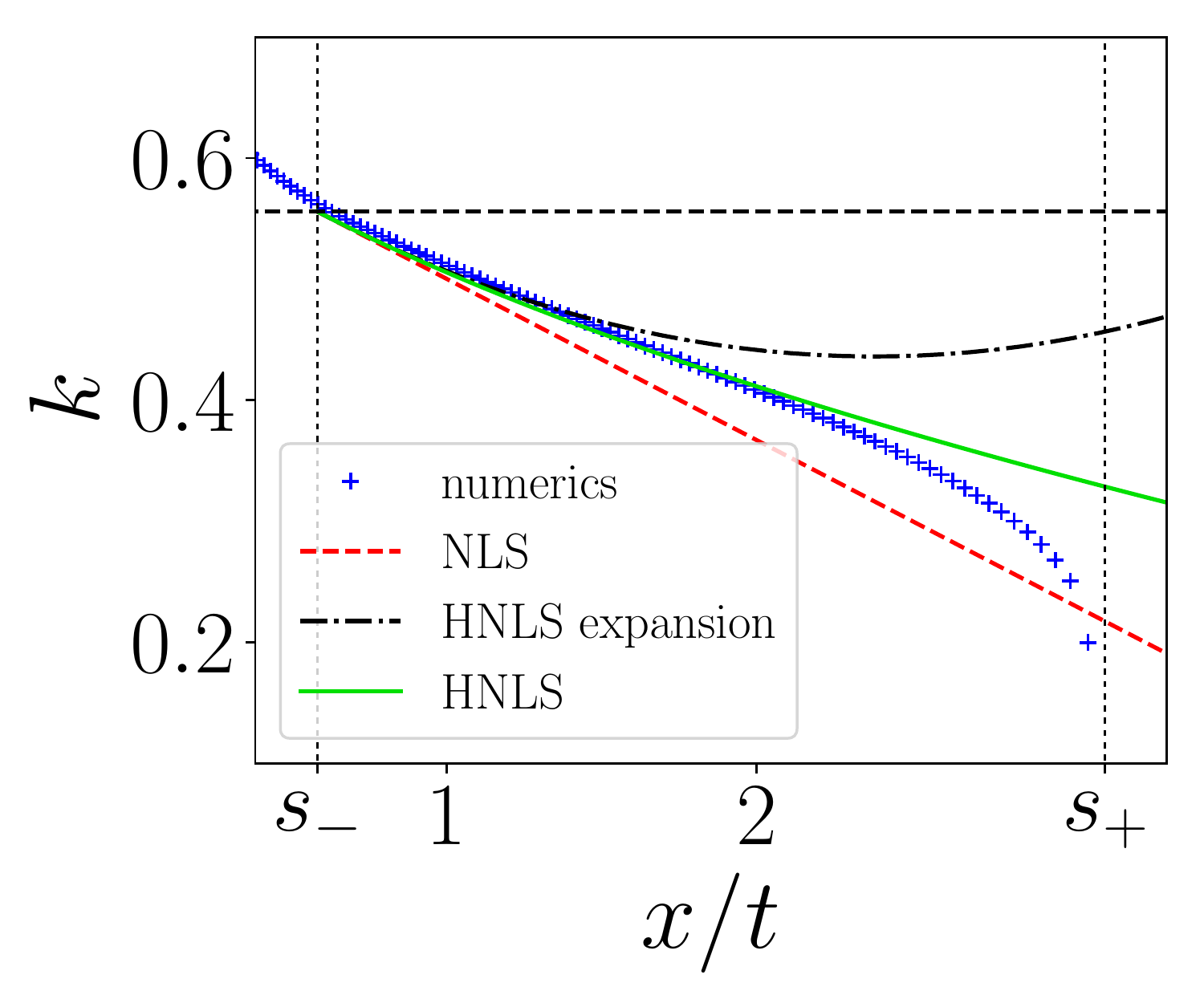}
%\end{subfigure}
%\begin{subfigure}[t]{.33\textwidth}
%\centering
\includegraphics[width=0.5\textwidth]{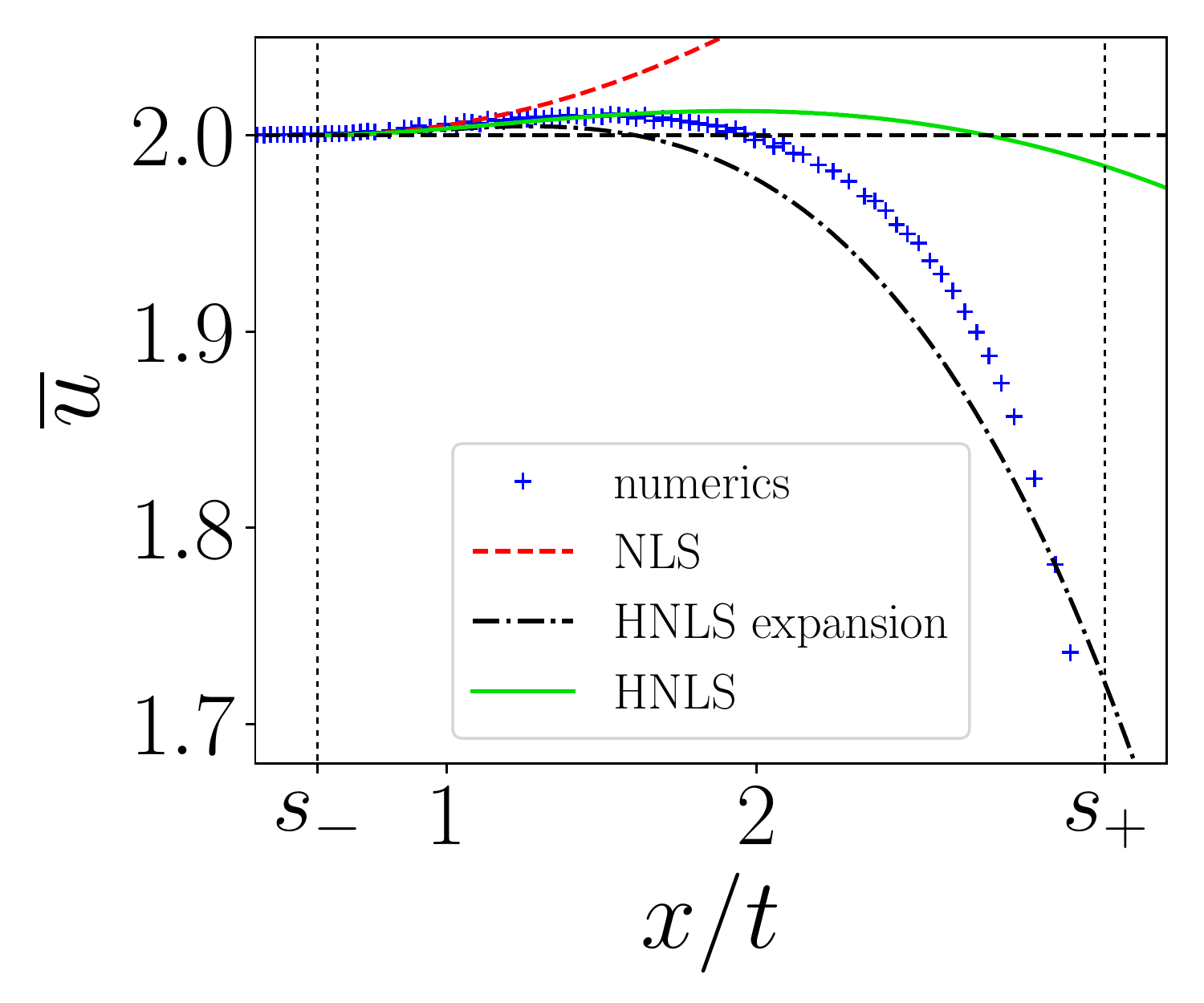}
%\end{subfigure}
\caption{Amplitude, wavenumber, and mean field profiles for the
  conduit Riemann problem with $(u_-,u_+)=(2,1)$ at $t=500$ (blue
  pluses), the NLS/HNLS asymptotic descriptions given by (i)
  expansions \eqref{dexplicit} with the first order (NLS)
  approximation (dashed red line) and the second order (HNLS)
  approximation (dash-dotted black line) and (ii) the exact simple wave
  solution Eqs.~\eqref{Vd_sim}, \eqref{ODE_HNLS} of the dispersionless
  HNLS equation with coefficients~\eqref{conduit_coef},
  \eqref{conduit_coef2} (solid green line). The horizontal black
  dashed lines correspond to the values of the corresponding fields in
  the harmonic limit: $a=0$, $k=k_- \simeq 0.56$ and $\ubar = u_-=2$.}
\label{fig:compare_conduit}
\end{figure}
The comparisons between the dispersionless HNLS vacuum rarefaction
simple wave solution \eqref{Vd_sim}, \eqref{ODE_HNLS} with
coefficients \eqref{conduit_coef}, \eqref{conduit_coef2}, its
asymptotic expansions \eqref{mod_general_harmonic1}, \eqref{dexplicit}
and the numerical solution of the Riemann problem for the conduit
equation are shown in Fig.~\ref{fig:compare_conduit}. As in the
previous cases, the full simple wave solution exhibits very good
agreement with the direct numerical solution over a broad DSW region,
while the first- and second-order approximations work satisfactorily
only in a relatively narrow vicinity of the harmonic edge.
%\MH{Recommend trying simulation with initial step width an order of
%  magnitude larger ($\xi = 20$ rather than $\xi = 2$).  Maybe the
%  discrepancy will be less.}

%We note in conclusion that it is implicit in the expansions \eqref{}  that $|\beta|=O(1)$  which is the case for equations with convex dispersion like the KdV and Serre equations.   in the conduit equation case one should take care that , i.e. one is not too close to the zero dispersion point 

%If $u_+$ is sufficiently small, $k_- < \sqrt{3 u_-}$
%(cf. Fig.~\ref{fig:compare_conduit}) and the NLS equation describing
%the DSW modulation close to the trailing edge enters the so-called
%focusing regime ($-\beta(k_-,u_-) \gamma(k_-,u_-)<0$ ,
%cf. Sec.~\ref{sec:small_ampl_dsw}): in this regime the envelop of the
%DSW is modulationally instable. Both defocusing-and-focusing regimes
%are displayed in Fig.~\ref{fig:compare_conduit}.

% The comparison between the asymptotic expansion~\eqref{dexplicit} with
% coefficients~\eqref{} and the direct numerical solution of the GNLS
% Riemann problem is displayed in Fig.~\ref{fig:GNLS}. In this scenario,
% the NLS description is not precise enough to describe the variation of
% the amplitude along the whole DSW and it becomes necessary to consider
% higher order nonlinear terms in the asymptotic expansion.

\newpage
\subsection{Serre equations}
\label{sec:serre}

The presentation until now has emphasized scalar dispersive
hydrodynamic equations in the form \eqref{uni_dh}.  Our methodology,
however, can be applied to systems of dispersive hydrodynamic
equations.  As an example, we now consider the Serre system modeling
fully nonlinear shallow water waves \cite{serre_contribution_1953,
  su_kortewegvries_1969}
\begin{equation}
\begin{split}\label{serre}
&\eta_t + (\eta u)_x = 0 ,\\
&u_t + u u_x + \eta_x = \frac{1}{\eta} \left(\frac{\eta^3}{3} \left[
u_{xt} + u u_{xx} - (u_x)^2 \right] \right)_x\, .
\end{split}
\end{equation}
We defer technical calculations to Appendices \ref{app:serre} and
\ref{app:HNLS_serre}.  Here $\eta$ is the total depth of the fluid and
$u$ is the depth averaged horizontal velocity.  The linear dispersion
relation of \eqref{serre} for small amplitude waves propagating on the
background $(\eta_0, u_0)$ has the form
\begin{equation}\label{ldr_serre}
\omega= \omega_0^{\pm}(k,u_0,\eta_0) = k\left( u_0 \pm \sqrt{\frac{\eta_0}{1+\eta_0^2
k^2/3}} \right).
\end{equation}
We assume Riemann initial data for \eqref{serre}
\begin{equation}
  \label{Riemann_Serre}
  \eta(x,0) =
  \begin{cases}
    \eta_- & x < 0 \\
    \eta_+ & x > 0
  \end{cases}, \quad u(x,0) =
  \begin{cases}
    u_- & x < 0 \\
    u_+ & x > 0
  \end{cases}
\end{equation}
 subject to  an additional constraint
\begin{equation}
  \label{serre_locus}
  u_+ - 2 \eta_+^{1/2} = u_- - 2 \eta_-^{1/2}
\end{equation}
that ensures a simple wave, 2-DSW resolution of \eqref{Riemann_Serre}
corresponding to the fast `$+$' mode in the dispersion relation
\eqref{ldr_serre} \cite{el_unsteady_2006, el_dispersive_2016}.  Due to
scaling and Galilean symmetries of \eqref{serre}, we can assume
$\eta_+=1$, $u_+=0$.

While the Serre system \eqref{serre} is not integrable, it satisfies
the pre-requisites of the DSW fitting method except for the loss of
genuine nonlinearity in a certain parameter regime, discussed further
below.  The corresponding analysis has been carried out in
\cite{el_unsteady_2006}, demonstrating excellent agreement with direct
numerical simulations. An implicit expression for the harmonic
(trailing) edge wavenumber $k_-$ of a 2-DSW as a function of the
Riemann data \eqref{Riemann_Serre}, obtained in \cite{el_unsteady_2006}
by integrating a bi-directional generalization of the ODE
\eqref{ode_char}, has the form
\begin{equation}
  \label{serre_k0}
  \begin{split}
    &\sqrt{\alpha} \Delta - \left(\frac{4-\alpha}{3}
    \right)^{21/10}\left(\frac{1+\alpha}{2} \right)^{2/5} =0,\\
    & \hbox{where} \quad \alpha
    = \left(\frac{2+s_-}{\sqrt \Delta} -2 \right)^{1/3}.
  \end{split}
\end{equation}
Here, $\Delta = \eta_- / \eta_+$ and
$s_-=\partial_k\omega_0^+(k_-,u_-,\eta_-)$. The result
\eqref{serre_k0} is valid as long as $\partial s_-/\partial \eta_- <
0$ (DSW fitting admissibility \cite{hoefer_shock_2014}) leading to the
condition $\Delta < \Delta_{c} \approx 1.43$ \cite{el_unsteady_2006}.
When $\Delta = \Delta_{c}$, the Whitham modulation system loses
genuine nonlinearity at the trailing edge. 
% \MH{Thibault: Can you
%  verify that one of \eqref{eq:1} or $\gamma \ne (\lambda-\mu)v$ is
%  violated at $\Delta = \Delta_c$?  If so, we should highlight this.
%  Loss of genuine nonlinearity could be viewed as a bifurcation in the
%  shock structure.}
%  \TC{
%Criteria $\gamma \ne (\lambda-\mu) v $ and \eqref{eq:1} can be rewritten
%in the following conditions for $\rho$ and $v$:
%\begin{equation}
%\begin{split}
%&\rho \neq \rho_c = -\frac{3(3\delta\gamma - \beta(\lambda-\mu))^2}{16
%\delta\mu( \lambda^2-\mu^2)}, \, v \neq v_{c,1} = \frac14 \left( \frac
%\beta \delta + \frac{\gamma}{\lambda-\mu} \right),\, \\
%&v \neq v_{c,2} =
%\frac{\gamma}{\lambda-\mu}.
%\end{split}
%\end{equation}
%  }

The multiple scales asymptotic expansion for the Serre equations
\eqref{serre} that lead to the NLS equation \eqref{NLS} are carried
out in Appendix~\ref{app:serre}. In these expansions, the envelopes of
the small amplitude oscillations of $\eta(x,t)$ and $u(x,t)$ are
proportional to each other: 
\begin{equation}
\label{serre_prop}
\begin{split}
&\begin{pmatrix}
\eta(x,t)\\u(x,t)
\end{pmatrix}=
\begin{pmatrix}
\eta_0 \\ u_0
\end{pmatrix} +
\left[\begin{pmatrix}
\tilde A(x,t) \\ \tilde B(x,t) 
 \end{pmatrix}e^{i[kx-\omega_0^+(k,\eta_0,u_0) t]} + \cc \right],\\
 &\text{where} \quad  \tilde B(x,t) = \tilde A(x,t) /
 \sqrt{\eta_0+\eta_0^3 k^2 /3}.
\end{split}
\end{equation}
The coefficients of the NLS equation \eqref{NLS} for the envelope
$\tilde A(x,t)$ are found to be:
\begin{equation}\label{serre_coef}
  \begin{split}
    &\beta =-\frac{\eta_0^{3/2} \kappa}{2(1+\kappa^2/3)^{5/2}} ,\\
    &\gamma = \frac{243 + 297\kappa^2 + 42 \kappa^4 +\kappa^6
      + \kappa^8}{4\eta_0^{5/2} \kappa ( 27+9\kappa^2+\kappa^4)
      (1+\kappa^2 /3)^{3/2}} ,\\
    &b_{1,\eta}= -\frac{(9-\kappa^2)(3+\kappa^2)^3}{\eta_0
      \kappa^2(27+9\kappa^2+\kappa^4)} ,\\
    &b_{1,u}=-\frac{(3+2\kappa^2)(27+6\kappa^2+\kappa^4)}{\eta_0^{3/2}\kappa^2
      \sqrt{1+\kappa^2/3} (27+9 \kappa^2+\kappa^4)} ,
  \end{split}
\end{equation}
where $\kappa = \eta_0 k$.  Here $b_{1,\eta}, b_{1,u}$ are the
coefficients in the mean flow expansions $\overline{\eta} = \eta_0 +
b_{1,\eta} \rho$ and $\ubar = u_0 + b_{1,u} \rho$.  The NLS
equation for the component $\tilde B$ is obtained by combining the NLS
equation for $\tilde A$ and the proportionality
relation~\eqref{serre_prop}.

\begin{figure}
\includegraphics[width=0.5\textwidth]{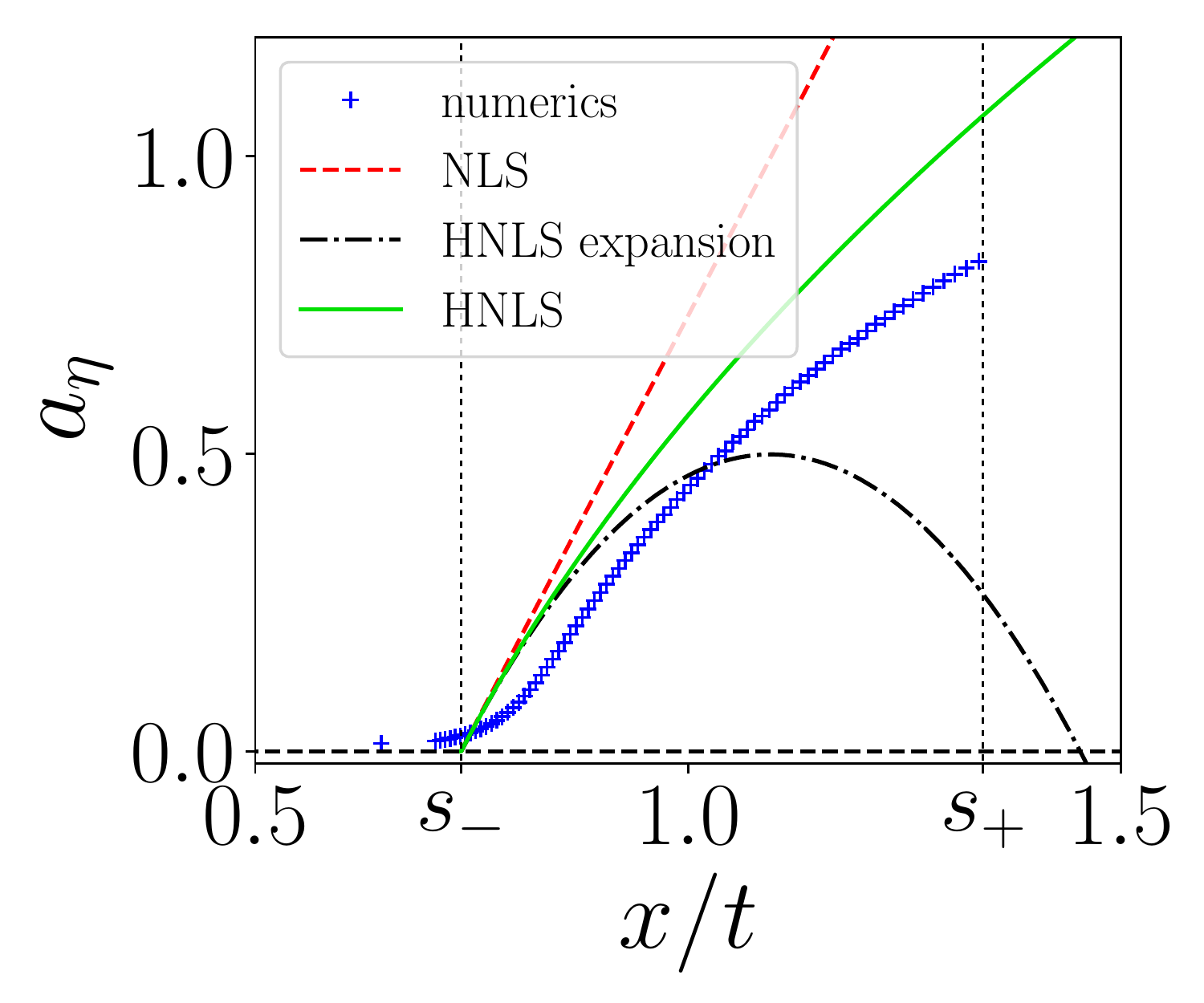}
\includegraphics[width=0.5\textwidth]{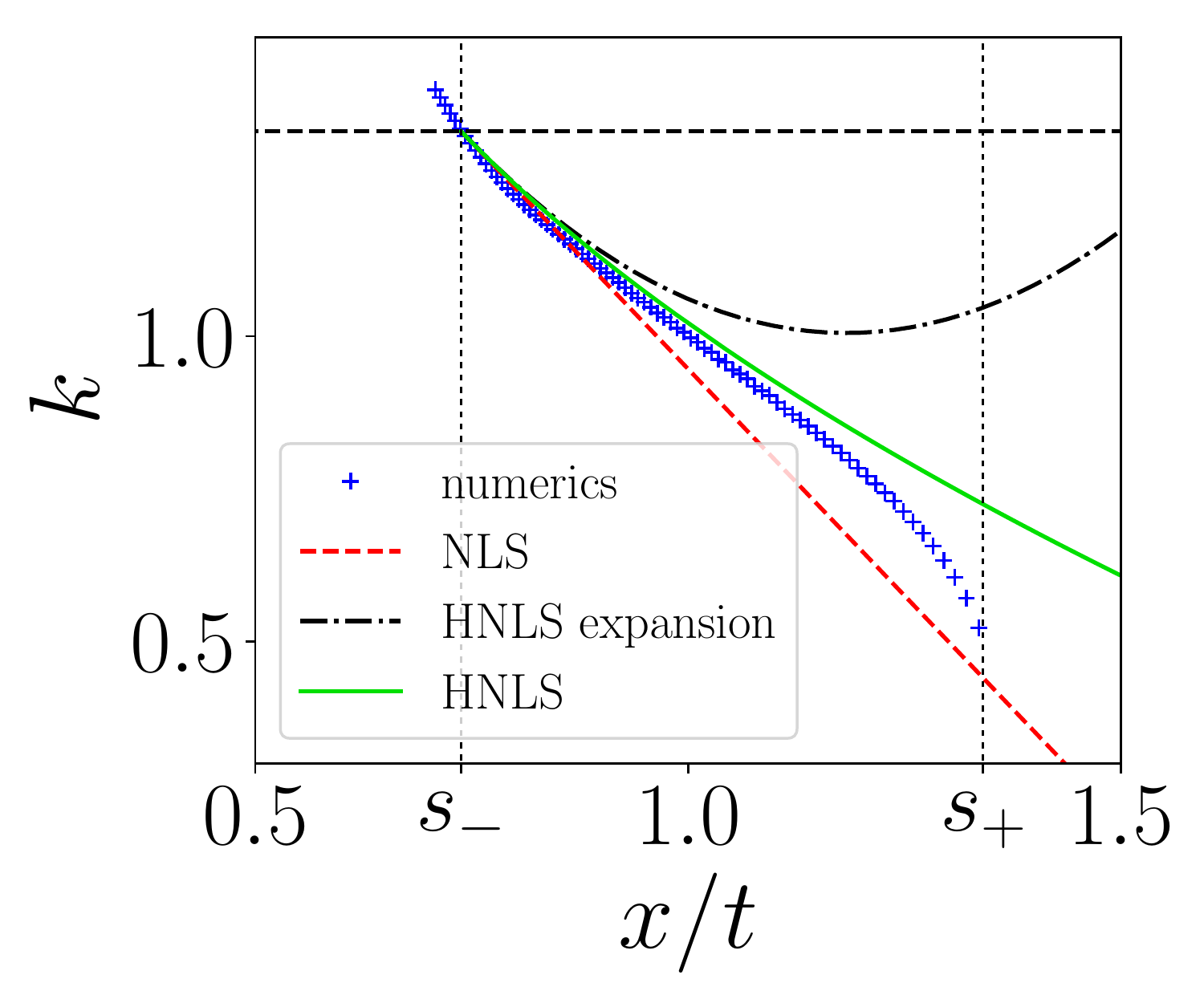}
\includegraphics[width=0.5\textwidth]{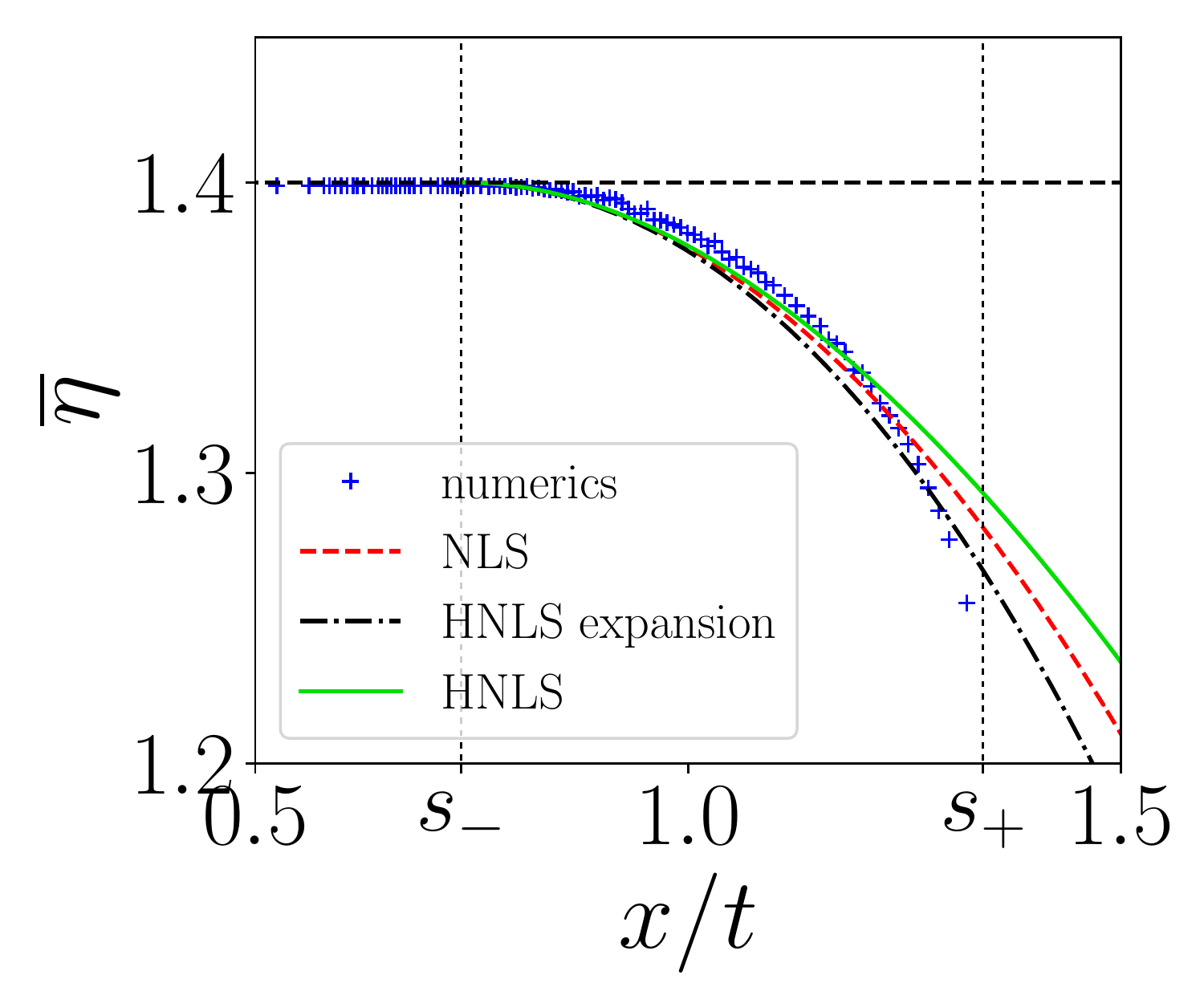}

\caption{Amplitude, wavenumber, and mean field profiles for the Serre
  Riemann problem with $(\eta_-,\eta_+)=(1.4,1)$ and
  $(u_-,u_+)=(2\sqrt \eta_--2,0)$.  Comparison between direct
  numerical simulations at $t=800$ (blue pluses), the NLS/HNLS
  asymptotic descriptions given by (i) expansions \eqref{dexplicit}
  with the first order (NLS) approximation (dashed red line) and the
  second order (HNLS) approximation (dash-dotted black line) and (ii) the
  exact simple wave solution Eqs.~\eqref{Vd_sim}, \eqref{ODE_HNLS} of
  the dispersionless HNLS equation with
  coefficients~\eqref{serre_coef}, \eqref{serre_coef2}
  and~\eqref{serre_coef3} (solid green line). The horizontal black
  dashed lines correspond to the values of the corresponding fields in
  the harmonic limit: $a_\eta=0$, $k=k_- \simeq 1.34$ and
  $\overline{\eta} = \eta_-=1.4$.}
\label{fig:serre}
\end{figure}

Going to $O(\eps^4)$, we obtain the coefficients $\delta, \lambda,
\mu$ in the HNLS equation \eqref{HNLS}, as well as the coefficients
$b_{2,\eta}, b_{2,u}$ in the second order expansions of the mean flow
$\overline{\eta}$, $\ubar$ in terms of $k, u_0, \eta_0$. These are
presented in Appendix \ref{app:HNLS_serre} (see formulae
\eqref{serre_coef2} and~\eqref{serre_coef3}).  In this higher order
description, the envelopes $\tilde A$ and $\tilde B$ are no longer
proportional to each other, and the coefficient $\mu$ in the HNLS
equation has to be derived separately for each component.
% (cf. Sec.~\ref{app:hnls_multicomp}).

To apply the HNLS equation to the description of the vicinity of the
Serre DSW harmonic edge, we set $u_0=u_-$, $\eta_0=\eta_-$, $k=k_-$,
where the dependence of $k_-$ on the Riemann data
\eqref{Riemann_Serre} is obtained numerically from the implicit
equation \eqref{serre_k0}.

The comparison between the simple wave solution
\eqref{Vd_sim},\eqref{ODE_HNLS} with parameters given by
\eqref{serre_coef}, \eqref{serre_coef2}, \eqref{serre_coef3} and the
numerical solution of the Riemann problem for the Serre equations is
displayed in Fig.~\ref{fig:serre}. Also shown are the curves
corresponding to the first order \eqref{mod_general_harmonic1} and the
second order \eqref{dexplicit} approximations of the full solution
\eqref{Vd_sim},\eqref{ODE_HNLS}. The comparisons are made for the DSW
amplitude $a$, the wavenumber $k$ and the mean depth
$\overline{\eta}$. We can see that the full simple wave solution
\eqref{Vd_sim},\eqref{ODE_HNLS} of the dispersionless HNLS equation
provides a more accurate description of the DSW modulation than the
first and second order approximations.  Similar to the KdV case, the
full simple wave solution of the dispersionless HNLS equation provides
a good approximation of the nonlinear wave modulation over a
significant portion of the DSW, well beyond the formal applicability
of the small amplitude (H)NLS approximation. On the other hand, we can
see that, in contrast to the KdV case, the second order approximation
\eqref{dexplicit} develops quite strong deviation from the actual
modulation for moderate values of $(x/t - s_-)$.

The final comment concerns the already discussed generic discrepancy
between the simple wave modulation DSW solution and direct numerical
solution of the Riemann problem in the vicinity of the harmonic edge
(see the discussion at the end of Sec.~\ref{sec:kdv}). This
discrepancy is more pronounced for the Serre equations than for the
KdV equation although the overall agreement with the dispersionless
HNLS solution is still quite good.  

\section{Conclusion and Discussion}
\label{sec:discussion}

In this work, we have developed an efficient, universal approach for
the analytical description of the interior  structure of a
dispersive shock wave (DSW) that extends the previously developed DSW
fitting method \cite{el_resolution_2005} for the DSW edge speeds.  The
key element of the  extension is the realization that the DSW
modulation described by an expansion fan solution of the Whitham
modulation equations can be universally approximated, in the vicinity
of the weakly nonlinear harmonic edge, by a special vacuum rarefaction
solution of the shallow water equations. The connection between the
original dispersive hydrodynamics and the approximating shallow water
system occurs via a long-wave, dispersionless limit of the NLS
equation for weakly nonlinear, narrow-band Stokes waves, whose
parameters are determined by DSW fitting when the NLS equation is of defocusing type.
The NLS
type (defocusing or focusing) determines DSW stability properties. The
developed approach is particularly attractive for applications as it
allows one to avoid a potentially complex, full Whitham modulation
analysis of DSWs in favor of the more straighforward and standard NLS
theory.

 The efficacy of the
developed approach is demonstrated by several representative examples
including the KdV equation, the Serre shallow water equations and the
viscous fluid conduit equation, the two latter systems being
non-integrable. In all considered cases, it is shown that the
inclusion of higher order terms in the NLS equation dramatically
improves agreement between the approximate modulation solution and the
numerical solution of the original dispersive Riemann problem.  
The proposed method has broad implications for DSW analysis in
non-integrable systems, where exact methods based on the inverse
scattering theory are not available. One interesting perspective is to
use the NLS approximation for the analytical description of multiphase
 modulations that are symptomatic of  DSW implosions  (see
\cite{lowman_dispersive_2013} and Sec.~\ref{sec:conduit} of this
paper).  In this context, the description will necessarily depend upon
dispersive terms in the NLS equation, which do not play a role in the
classical expansion fan DSW solutions considered in the present paper.
Further, the improved DSW description in
Sec.~\ref{sec:higher_order_nls},  based on the higher order NLS 
equation,  provides a general mathematical
framework for DSW analysis in systems with non-convex dispersion,
which are currently under active investigation
\cite{el_radiating_2016, sprenger_shock_2017}.  We also envisage
intriguing connections with the multisymplectic theory of universal
dispersive deformations of the Whitham equations near coalescing
characteristics,  precisely the configuration that occurs at
the DSW harmonic edge; see \cite{ratliff_whitham_2016,
  bridges_symmetry_2017} and references therein.

Probably the most appealing extension of the developed harmonic edge
DSW structure theory would be to find its counterpart in the
vicinity of the DSW soliton edge (utilizing small $k$ asymptotics), and
to construct a universal, matched, uniformly valid asymptotic solution for
the entire DSW modulation.

% Acknowledgements
\section*{Acknowledgments}

The research of TC and GAE was supported by EPSRC grant
EP/R00515X/1. The research of MS and MAH was supported by National
Science Foundation grants: DMS-1517291 and DMS-1812445 (MS), and CAREER DMS-1255422 and 
DMS-1816934 (MAH). The authors acknowledge useful discussions with Dan
Ratliff.

\begin{appendix}

\section{Derivation of the HNLS equation for the  conduit equation}
\label{app:HNLS_conduit}

The derivation of the HNLS equation for KdV~\eqref{kdv} is detailed in
Ref.~\cite{boyd_weakly_2001} and we simply repeat in this section the
main steps of the derivation which are applicable to the conduit
equation ~\eqref{conduit}.  In order to derive the HNLS
equation~\eqref{HNLS}, we look for the solution of \eqref{conduit} in
the form
\begin{equation}
u = u_0+\eps u_1+\eps^2 u_2+\eps^3 u_3+\dots
\end{equation}
with 
$$
u_1=A(X,T_1,T_2,T_3)e^{i(kx-\omega t)}+c.c,  \ X=\eps x, T_1=\eps t,
T_2=\eps^2 t, T_3=\eps^3 t.
$$
The cancellation of the secular terms at $O(\eps^2)$ and $O(\eps^3)$
in the expansion in $\eps$ gives  Eqs.~\eqref{vg}
and~\eqref{NLS} respectively, and the cancellation of the secular term at
$O(\eps^4)$ gives
\begin{equation}
\label{HNLSb}
A_{T_3} + \delta A_{XXX} + \lambda |\tilde A|^2 A_X + \mu A^2 A^*_X = 0,
\end{equation}
As a by-product of the $O(\eps^4)$ expansion we also obtain a higher order
correction of the mean $\ubar$ which reads:
\begin{equation}
\ubar - u_0 \simeq \eps^2 b_1 |A|^2 + \eps^3 b_2 i(A A_X^* - A^* A_X).
\end{equation}
We define the un-scaled envelope $\tilde A(x,t)$ by:
\begin{equation}
\label{unscaled}
\tilde A(x,t) = \eps A(\eps x, \eps t, \eps^2 t, \eps^3 t),
\end{equation}
implying the new substitution rule (cf. Eq.~\eqref{substitut}):
\begin{equation}
\label{substitut2}
\tilde A_t = \eps^2 A_{T_1} + \eps^3
A_{T_2}+ \eps^4 A_{T_3}.
\end{equation}
Combining Eqs.~\eqref{vg}, \eqref{NLS}, \eqref{HNLSb} with the
substitution rule~\eqref{substitut2}, we obtain the
HNLS equation~\eqref{HNLS} for the un-scaled envelope $\tilde A(x,t)$.

Applying the above algorithm to the conduit equation~\eqref{conduit} we find the coefficients of the HNLS equation \eqref{HNLS} to be:
\begin{equation}
\label{conduit_coef2}
\begin{split}
&\delta  = \frac{2u_0^2(1-6u_0k^2+u_0^2k^4)}{(1+u_0k^2)^4}, \\
&\lambda =
\frac{2(-9+9u_0k^2+21u_0^2k^4+23u_0^3k^6+8u_0^4k^8)}
{3u_0^2k^2(3+4u_0k^2+u_0^2k^4)^2},\\
&\mu =
\frac{-27+24u_0k^2+55u_0^2k^4+8u_0^3k^6}
{3u_0^2k^2(1+u_0k^2)(3+u_0k^2)^2},\\
&b_2 = \frac{2(3+2u_0k^2+7u_0^2k^4)}{u_0^2k^3(3+u_0k^2)^2}.
\end{split}
\end{equation}

\section{Derivation of the NLS equation for the Serre equations}
\label{app:serre}

We detail in this Appendix the multiple scales expansions for the
Serre equations \eqref{serre} leading to the NLS equation
\eqref{NLS}. Although the derivation of the NLS equation for
% multi-component
systems is standard (see, e.g., \cite{taniuti_perturbation_1969}), the
computation can be rather cumbersome because of the vectorial nature
of the system; this difficulty can be overcome using symbolic
computations, as we have done in this case (we used {\it Mathematica}).
Similar to the multiple scales asymptotic expansions for scalar
equations, we look for the solution in the form
\begin{equation}\label{serre_exp}
\Xi =
\begin{pmatrix}
\eta \\ u
\end{pmatrix}
=
\begin{pmatrix}
\eta_0\\u_0
\end{pmatrix}
+\eps \Xi_1+\eps^2 \Xi_2+\eps^3 \Xi_3+\dots
\end{equation}
with 
$$
\Xi_1=\Psi(X,T_1,T_2)e^{i(kx-\omega t)}+c.c,  \ X=\eps x, T_1=\eps t, T_2=\eps^2 t,
$$
where $\Psi \in \mathbb{C}^2$ is a complex two-component vector.
% The envelop $A$ of the wave packet is now a vector, and for the sake
% of clarity we denote its two components as $A$ and $B$.
Substituting \eqref{serre_exp} into the Serre system \eqref{serre} and collecting the 
$O(\eps)$ terms we get
\begin{equation}
\label{serre1bis}
M(ik,-i\omega)\Psi = 0,
\end{equation}
where 
\begin{equation}\label{serre1}
M(ik,-i\omega) =
\begin{pmatrix}
i(u_0k-\omega) & i \eta_0 k\\
i k &
i(1+\eta_0^2 k^2/3) (u_0 k -\omega)
\end{pmatrix}.
\end{equation}
The null space of \eqref{serre1} is not empty if
\begin{equation}\label{ldr_serre1}
\omega= \omega_0^{\pm}(k,u_0,\eta_0) = k\left( u_0 \pm \sqrt{\frac{\eta_0}{1+\eta_0^2
k^2/3}} \right).% \omega = \omega_0(k) = \frac{u_0}{\eta_0} \kappa +
% \sqrt{\frac{\eta_0}{1+\kappa^2/3}},
\end{equation}
% where $\kappa = \eta_0 k$ is a convenient parameter.
Eq.~\eqref{ldr_serre1} is nothing but the linear dispersion relation
\eqref{ldr_serre} of the Serre system.  The 2-DSW developing in the Riemann
problem~\eqref{serre},\eqref{Riemann_Serre},\eqref{serre_locus}
corresponds to the fast `+' mode (cf. Sec.~\ref{sec:serre} and
\cite{el_unsteady_2006, el_dispersive_2016}) so we assume in the
following that $\omega = \omega_0^+(k,\eta_0,u_0)$.
% which is the dispersion relation of the system.
Hence Eq.~\eqref{serre1bis} yields the non trivial solution:
\newcommand{\qa}{\sqrt{\eta_0(1+ \kappa^2 /3)}}
\begin{equation}
\label{Psi}
\Psi = \Psi^+_0(X,T_1,T_2)=
\begin{pmatrix}
1\\ 1/ \sqrt{\eta_0(1+ \kappa^2 /3)}
\end{pmatrix}
 \,A(X,T_1,T_2),
\end{equation}
where $\kappa = \eta_0 k$ is a
convenient parameter and $A(X,T_1,T_2)$ is now a scalar.
% In the following $A$ denotes the component $B$. 
% It is also useful to
% derive at this point t
The kernel of $M^T$ is spanned by $\{L_0^+, L^-\}$ where
\begin{equation}
L_0^\pm =
\left( \pm
\sqrt{1+ \kappa^2 /3},\,
\sqrt \eta_0 \right)^T\, .
\end{equation}
Assuming that $\Psi$ is given by~\eqref{Psi} and
$\omega=\omega_0^+(k,\eta_0,u_0)$, the second order of the asymptotic
expansion reads
\begin{equation}\label{serre2}
M(\partial_x,\partial_t) \Xi_2 = C_1  e^{i(kx - \omega_0^+t)} + C_2  e^{2i(kx -
\omega_0^+t)} + \cc ,
\end{equation}
where we drop the dependences of the dispersion relation $\omega_0^+$
by convenience. The vectors $C_1$ and $C_2$ are given by:
$$
C_1 =
-\begin{pmatrix}
\displaystyle
A_{T_1} + \left(u_0 + \sqrt{\frac{\eta_0}{1+ \kappa^2 /3}} \right)A_X
\\
\displaystyle
\sqrt{\frac{1+ \kappa^2 /3}{\eta_0}} A_{T_1} + \left(  u_0  \sqrt{\frac{\eta_0}{1+ \kappa^2 /3}} +
\frac{3- \kappa^2}{3+\kappa^2}\right)A_X
\end{pmatrix},
$$
and
$$
C_2=
-\begin{pmatrix}
\displaystyle
\frac{2i\kappa}{\eta_0^{3/2}\sqrt{1+ \kappa^2 /3}} \,A^2\\
\displaystyle
\frac{i\kappa(3-5\kappa^2)}{\eta_0^2(3+\kappa^2)} \,A^2
\end{pmatrix}.
$$
Since $\det M(ik,-i\omega_0^+)=0$, a compatibility condition is
necessary to solve Eq.~\eqref{serre2} (cf. for instance
Ref. \cite{taniuti_perturbation_1969}), thus we impose the ortogonality requirement
% In order to cancel For multi-component systems the ``cancellation'' of the secular term
% $\propto e^{i(k x -\omega t)}$ is the condition $C_1 \in (\ker
% \,^tM)^\perp$, see for instance~\cite{taniuti_reductive_1974}, which is in this case
$ L_0^+ \cdot C_1=0$.
This condition is satisfied if the wave packet propagates with the
group velocity~$\partial_k \omega_0^+(k,\eta_0,u_0)$:
\begin{equation}\label{serre2cond}
A_{T_1} +  \partial_k \omega_0^+(k,\eta_0,u_0) \, A_X = 0.
\end{equation}
Providing that~\eqref{serre2cond} is respected, one solution
of~\eqref{serre2} is:
\begin{equation}\label{serre2sol}
\begin{split}
\Xi_2 =
\begin{pmatrix}
Q \\ R
\end{pmatrix}
&+
\left[\begin{pmatrix}
0 \\ \displaystyle\frac{i\sqrt{3\eta_0} \kappa}{(3+\kappa^2)^{3/2}}
\end{pmatrix}A_X e^{i(kx - \omega_0^+t)} + \cc \right]\\
&+
\left[
\begin{pmatrix}
\displaystyle
\frac{3+\kappa^2}{2\eta_0 \kappa^2}\\
\displaystyle
\frac{3-\kappa^2}{2\eta_0^{3/2} \kappa^2 \sqrt{1+\kappa^2/3}}
\end{pmatrix}A^2e^{2i(kx - \omega_0^+t)} +\cc \right],
\end{split}
\end{equation}
where $Q(X,T_1,T_2)$ and $R(X,T_1,T_2)$ are two unknown fields that
remain to be determined. $Q$ and $R$ are necessary for the
consistency of the asymptotic expansion, as we shall see at the next
order (cf. Eqs.~\eqref{QR}).

The solution~\eqref{serre2sol} is not unique and the general solution
of~\eqref{serre2} reads as:
$\Xi_2 + \left[ K(X,T_1,T_2) e^{i(kx - \omega_0^+t)} + \cc \right]$
where $K(X,T_1,T_2)$ belongs to the kernel of
$M(ik,-i\omega_0^+)$. This additional term does not modify the NLS
equation~\eqref{NLS} in the end but it plays an important role in
higher order descriptions (cf.~Appendix~\ref{app:HNLS_serre}). In
practice, we choose $K(X,T_1,T_2)$ such that one of the two components
of $\Xi_2$ proportional to $e^{i(kx - \omega_0^+t)} $ is equal to $0$
(which is already the case here).% ;
% note that generally, we cannot cancel both components.

Finally, if we substitute $\Xi_2$ by the solution~\eqref{serre2sol},
the $O(\eps^3)$ of the expansion reads
\begin{equation}\label{serre3}
M(\partial_x,\partial_t) \Xi_3 = D_0
+ [D_1  e^{i(kx - \omega_0^+t)} + D_2  e^{2i(kx -
\omega_0^+t)} + D_3  e^{3i(kx -
\omega_0^+t)}+ \cc] ,
\end{equation}
with
$$
D_0=
-\begin{pmatrix}
\displaystyle
Q_{T_1}+u_0 Q_X + \eta_0 R_X +\frac{(|A|^2)_X}{\qa}\\
\displaystyle
R_{T_1} + Q_X + u_0 R_X +
\frac{9+6\kappa^2-\kappa^4}{\eta_0(3+\kappa^2)^2} (|A|^2)_X
\end{pmatrix},
$$
and
\begin{equation*}
\begin{split}
D_1=\;&
-\begin{pmatrix}
\displaystyle
A_{T_2} + \frac{i\sqrt 3 \eta_0^{3/2} \kappa}{(3+\kappa^2)^{3/2}}
A_{XX}+
\frac{3i}{\eta_0^{5/2} \kappa \sqrt{1+\kappa^2/3}}|A|^2A
\\
\displaystyle
\sqrt{\frac{1+\kappa^2/3}{\eta_0}}A_{T_2} +
\frac{i\eta_0\kappa(6-\kappa^2)}{(3+\kappa^2)^2}
A_{XX} + \frac{i(9+7\kappa^4)}{2\eta_0^3 \kappa(3+\kappa^2)} |A|^2A
\end{pmatrix}\\
&-\begin{pmatrix}
\displaystyle
\frac{i\kappa}{\eta_0^{3/2} \sqrt{1+\kappa^2/3}}Q A +\frac{i
\kappa}{\eta_0}R A\\
\displaystyle
\frac{-2i \kappa^3}{\eta_0^2(3+\kappa^2)}Q A +\frac{i\kappa
\sqrt{1+\kappa^2/3}}{\eta_0^{3/2}} R A 
\end{pmatrix}.
\end{split}
\end{equation*}
We do not present the coefficients $D_2$ and $D_3$ for the second and
third harmonic terms at this stage since they are not needed for the
derivation of the NLS equation. However, these terms are needed to
solve~\eqref{serre3}, and ultimately derive the HNLS equation at the
next order.

Since $M(0,0)=0$, the constant term $D_0$ should be equal to $0$. This
condition is respected for $Q(X,T_1,T_2)$ and $R(X,T_1,T_2)$ given by:
\begin{equation}\label{QR}
\begin{split}
&Q= -\frac{(9-\kappa^2)(3+\kappa^2)^3}{\eta_0
\kappa^2(27+9\kappa^2+\kappa^4)} |A|^2,\\ 
&R= -\frac{(3+2\kappa^2)(27+6\kappa^2+\kappa^4)}{\eta_0^{3/2}\kappa^2
\sqrt{1+\kappa^2/3} (27+9 \kappa^2+\kappa^4)} |A|^2.
\end{split}
\end{equation}
Providing that $Q$ and
$R$ are substituted by the solutions~\eqref{QR}, the compatibility
condition $L_0^+ \cdot D_1=0$ gives
\begin{equation}
i A_{T_2} + \beta(k) A_{XX} + \gamma(k) |A|^2A = 0,
\end{equation}
with 
$$
\beta =-\frac{\eta_0^{3/2} \kappa}{2(1+\kappa^2/3)^{5/2}} ,\;
\gamma = \frac{243 + 297\kappa^2 + 42 \kappa^4 +\kappa^6
+ \kappa^8}{4\eta_0^{5/2} \kappa ( 27+9\kappa^2+\kappa^4)
(1+\kappa^2 /3)^{3/2}}.
$$

\section{Derivation of the HNLS equation for the Serre equations}
\label{app:HNLS_serre}

% \subsection{Multi-component systems}
% \label{app:hnls_multicomp}

The definition of the un-scaled envelope~\eqref{unscaled} is not always
adequate for systems.  Unlike scalar equations  where the only term of the solution $u(x,t)$
proportional to the first harmonic $e^{i(kx-\omega t)}$ is
$\eps A(X,T_1,T_2)$, solutions of ``multi-component''
systems (Eq.~\eqref{serre} for instance) % and~\eqref{GNLS}
can contain higher order
corrections of the envelope. % Indeed, in this
% higher order description one has to take into account potential
% correction $O(\eps^2)$ of the envelop
% (cf. Ref.~\cite{sedletsky_fourth-order_2003}).

For the Serre system~\eqref{serre} we define
(cf. Appendix.~\ref{app:serre}):
\begin{equation}
\begin{pmatrix}
\eta(x,t) \\ u(x,t)
\end{pmatrix} =
\begin{pmatrix}
\eta_0 \\ u_0
\end{pmatrix}+
\left[\begin{pmatrix}
\tilde A(x,t) \\ \tilde B(x,t)
\end{pmatrix}e^{i(kx - \omega_0^+t)} + \cc \right].
\end{equation}
We have shown in the previous section, cf. Eqs.~\eqref{Psi} and~\eqref{serre2sol},
that
\begin{equation}
\label{BB}
\begin{pmatrix}
\tilde A(x,t) \\ \tilde B(x,t)
\end{pmatrix} =
\begin{pmatrix}
\eps A(X,T_1,T_2,T_3) \\
\eps \alpha A(X,T_1,T_2,T_3) + i \eps^2 \sigma A_X(X,T_1,T_2,T_3) 
\end{pmatrix},
\end{equation}
with $\alpha = 1/ \sqrt{\eta_0(1+ \eta_0^2k^2 /3)}$ and
$\sigma=\sqrt{3\eta_0} \eta_0 k/(3+\eta_0^2
k^2)^{3/2}$. Definition~\eqref{unscaled} and the substitution
rule~\eqref{substitut2} still hold for the component
$\tilde A$ for which we obtain Eq.~\eqref{HNLS}. However, a new substitution rule
is necessary to derive the modulation equation for the ``total
amplitude''  $\tilde B$:
\begin{equation}
\tilde B_t = \eps^2 \alpha A_{T_1} + \eps^3
\alpha A_{T_2}+ \eps^4 \alpha A_{T_3} + i \eps^3 \alpha A_{X T_1} + i \eps^4
\alpha A_{X T_2}.
\end{equation}
% but no longer for the the component
% $\tilde B$. 
A careful derivation gives (cf. Ref.~\cite{sedletsky_fourth-order_2003}):
% \begin{equation}
% i \tilde A_t + i \omega_0'(k) A_x+  \beta  \tilde A_{xx} + \gamma | \tilde A|^2
% \tilde A  + i \delta
% \tilde A_{xxx} + i \lambda  |A|^2
% \tilde A_x+ i \mu  \tilde A^2  \tilde A^*_x= 0.
% \end{equation}
% As it was
% detailed in Ref.~\cite{sedletsky_fourth-order_2003}, one has to
% consider correction of $O(\eps^2)$ to derive the HNLS
% equation~\eqref{HNLS} for the ``total amplitude''. 
% Similarly, one can obtain the following HNLS equation for $\tilde B$
% (cf. Ref.~\cite{sedletsky_fourth-order_2003}:
\begin{equation}
\label{eq:B}
\begin{split}
&i \tilde B_t + i \omega_0'(k) B_x+  \beta  \tilde B_{xx} + \frac{\gamma}{\alpha^2} | \tilde B|^2
\tilde B  + i \delta
\tilde B_{xxx} + i \frac{\lambda}{\alpha^2}  |B|^2
\tilde B_x\\
&+ i \left(\frac{\mu}{\alpha^2}
+\frac{2\gamma
\sigma}{\alpha^3} \right)  \tilde B^2  \tilde B^*_x= 0.
\end{split}
\end{equation}
We notice that the coefficient in front of $ \tilde B^2  \tilde
B^*_x$ in~\eqref{eq:B} is not proportional to $\mu$ as one might expect
if one considered the inadequate definition $\tilde B = \eps \alpha
A$. Nonetheless, to the first order both $\tilde B = \eps \alpha
A$ and definition~\eqref{BB} yield the same NLS equation for $\tilde B$
which can be simply obtained by substituting in~\eqref{NLS3} $\tilde A$ by
$\tilde B/\alpha$.

We now present, without derivation, the coefficients for the HNLS
equation \eqref{HNLS} describing the envelope $\tilde A$ for the
component $\eta(x,t)$ of the Serre system. The envelope $\tilde B$ of
the component $u(x,t)$ can be put in the form~\eqref{BB} allowing for
the determination the corresponding coefficient $\sigma$. The
coefficients computed using {\it Mathematica} are:
\begin{equation}
\label{serre_coef2}
\begin{split}
&\delta  = \frac{3\sqrt 3 \eta_0^{5/2} (3-4\kappa^2)}{2(3+\kappa^2)^{7/2}},\\
&
\begin{split}
\lambda= \;&(9-\kappa^2) (-2187-3159\kappa^2 -567 \kappa^4+243\kappa^6
+ 90\kappa^8 \\ &+8\kappa^{10} ) / ( 6
\eta_0^{3/2}\kappa^2(1+\kappa^2/3)^{5/2} (27+9\kappa^2 + \kappa^4)^2
),
\end{split}\\
&
\begin{split}
\mu = \;&(-19683-24786\kappa^2 -11502 \kappa^4-2754\kappa^6 -276
\kappa^8 +14 \kappa^{10} \\ &+3\kappa^{12} ) /
(4\eta_0^{3/2}\kappa^2(1+\kappa^2/3)^{5/2} (27+9\kappa^2+\kappa^4)^2),
\end{split}\\
&\sigma = \frac{\sqrt{3\eta_0} \kappa}{(3+\kappa^2)^{3/2}},
\end{split}
\end{equation}
\begin{equation}
\label{serre_coef3}
\begin{split}
&b_{2,\eta}=
\frac{6(3+\kappa^2)(243+81\kappa^2+36\kappa^4+4\kappa^6)}{\kappa^3
(27+9\kappa^2 + \kappa^4)^2},\\
&
\begin{split}
b_{2,u}= &\sqrt 3 (13122+15309 \kappa^2+8505\kappa^4+2322
\kappa^6 +315 \kappa^8 +27 \kappa^{10} \\ &+ 2 \kappa^{12}) /
(\sqrt \eta_0 \kappa^3 (3+\kappa^2)^{3/2}(27+9\kappa^2 + \kappa^4)^2 ),
\end{split}
\end{split}
\end{equation}
where $\kappa = \eta_0 k$.

\section{Numerical Methods}
\label{app:num}

The initial step~\eqref{eq:riemann_data1} of the Riemann problem is
implemented numerically by the function:
\begin{equation}
\label{tanh}
u(x,t=0) = \frac{u_+-u_-}{2} \tanh \left( \frac{x}{\xi} \right) +
\frac{u_++u_-}{2},
\end{equation}
for the KdV ~\eqref{kdv} and the conduit ~\eqref{conduit} equations.  We implement a similar
step for the field $\eta(x,t)$ in the Serre system~\eqref{serre}. We shall use $\xi = 2$ in
our examples.
% for the different systems. The generation of linear excitations
% close to the harmonic edge of the DSW is unavoidable numerically and
% the width $\xi$ has been chosen sufficiently large to minimize their
% generation. These excitations do not enter the ``composition'' of
% the DSW and are not described by the DSW fitting method presented in
% this work. Cf. \cite{mesh} for the origin of these excitations
% {\color{red} (more recent papers?)}.
We choose periodic boundary conditions: $u(x +L) =u(x)$ (and
$\eta(x+L)=\eta(x)$); in practice we consider a domain $[0;L]$
sufficiently
large to avoid interactions with the boundaries.\\

% for the resolution of the Serre
% system and periodic boundary conditions for the other
% systems (in that case the initial condition~\eqref{tanh} is
% periodized). The type of boundary conditions has no incidence on the
% formation of the DSW since far from
% the initial step we have: $u_x(\pm \infty) = 0$.\\
We used a spectral method to solve the Riemann problem for the KdV
equation (cf. for instance Ref.~\cite{trefethen_spectral_2001}): we
rewrite Eq.~\eqref{kdv} in the form
\begin{equation}
\label{kdv_spec}
(e^{-ik^3 t} \hat u)_t + \, \frac{ik}{2} \widehat{u^2} = 0,
\end{equation}
where $\hat u$ and $\widehat{u^2} $ are spatial Fourier transforms of
$u$ and $u^2$ respectively.  The time integration of~\eqref{kdv_spec}
is performed through the 4th order explicit Runge-Kutta method and in
order to diminish the aliasing error we consider the ``Orszag 2/3 rule'' \cite{orszag_comparison_1972}.\\
% let $\hat v (k,t) = $ For the KdV and GNLS
% systems % explicit in time (KdV and GNLS)\footnote{The Riemann
% % problem for the GNLS system has also been solved using the Crank
% % --
% % Nicolson method. Both methods gave the same result for the
% % considered
% % periods of time.},
% the space derivatives are approximated using
% centered finite differences and the time
% integration is performed through the 4th order explicit Runge-Kutta method.\\

In order to solve the Riemann problem for the conduit equation, we
rewrite \eqref{conduit} in the following form (cf. for
instance Ref.~\cite{maiden_modulations_2016}):
\begin{align}
\label{wconduit}
&uw + 2 uu_x - (u^2 w_x)_x = 0,\\
\label{uconduit}
&u_t = w u.
\end{align}
Derivatives in~\eqref{wconduit} are approximated using centered finite
differences; then $w(x,t)$ is obtained by inverting the corresponding
banded linear system. \eqref{uconduit} is integrated through the 4th
order explicit Runge-Kutta method.

Similarly we rewrite the Serre system~\eqref{conduit} in the form:
\begin{align}
&w + u u_x + \eta_x = \frac{1}{\eta} \left(\frac{\eta^3}{3} \left[
w_x + u u_{xx} - (u_x)^2 \right] \right)_x,\\
&\eta_t + (\eta u)_x = 0 ,\quad u_t  = w,
\end{align}
and apply the same algorithm.

\end{appendix}

\bibliographystyle{abbrv}
%\bibliography{DSW_book}

\end{document}